\RequirePackage{fix-cm}
\documentclass[twocolumn,epjc3]{svjour3}  

\RequirePackage{graphicx}
\RequirePackage{mathptmx}      
\RequirePackage[numbers,sort&compress]{natbib}
\RequirePackage{amssymb} 
\RequirePackage{amsmath}
\RequirePackage{hyphenat}
\usepackage{rotating}

\smartqed  

\renewcommand{\vec}[1]{\boldsymbol{#1}}

\journalname{Eur. Phys. J. C}

\begin{document}

\title{All-flavour Search for Neutrinos from Dark Matter Annihilations in the Milky Way with IceCube/DeepCore}

\onecolumn
\author{IceCube Collaboration: M.~G.~Aartsen\thanksref{Adelaide}
\and K.~Abraham\thanksref{Munich}
\and M.~Ackermann\thanksref{Zeuthen}
\and J.~Adams\thanksref{Christchurch}
\and J.~A.~Aguilar\thanksref{BrusselsLibre}
\and M.~Ahlers\thanksref{MadisonPAC}
\and M.~Ahrens\thanksref{StockholmOKC}
\and D.~Altmann\thanksref{Erlangen}
\and K.~Andeen\thanksref{Marquette}
\and T.~Anderson\thanksref{PennPhys}
% ...
}
\authorrunning{IceCube Collaboration}
\thankstext{email}{Corresponding author: cph@physics.uu.se}
\thankstext{a}{Earthquake Research Institute, University of Tokyo, Bunkyo, Tokyo 113-0032, Japan}
\thankstext{b}{NASA Goddard Space Flight Center, Greenbelt, MD 20771, USA}
\institute{
% ...
Department of Physics, University of Adelaide, Adelaide, 5005, Australia \label{Adelaide}
% ...
\and Universit\'e Libre de Bruxelles, Science Faculty CP230, B-1050 Brussels, Belgium \label{BrusselsLibre}
% ...
\and Dept.~of Physics and Astronomy, University of Canterbury, Private Bag 4800, Christchurch, New Zealand \label{Christchurch}
% ...
\and Erlangen Centre for Astroparticle Physics, Friedrich-Alexander-Universit\"at Erlangen-N\"urnberg, D-91058 Erlangen, Germany \label{Erlangen}
% ...
\and Dept.~of Physics and Wisconsin IceCube Particle Astrophysics Center, University of Wisconsin, Madison, WI 53706, USA \label{MadisonPAC}
% ...
\and Department of Physics, Marquette University, Milwaukee, WI, 53201, USA \label{Marquette}
% ...
\and Physik-department, Technische Universit\"at M\"unchen, D-85748 Garching, Germany \label{Munich}
% ...
\and Oskar Klein Centre and Dept.~of Physics, Stockholm University, SE-10691 Stockholm, Sweden \label{StockholmOKC}
% ...
\and Dept.~of Physics, Pennsylvania State University, University Park, PA 16802, USA \label{PennPhys}
% ...
\and DESY, D-15735 Zeuthen, Germany \label{Zeuthen}
} 

\date{Received: date / Accepted: date}

\maketitle
\twocolumn

\begin{abstract}

 We present the first IceCube search for a signal of dark matter annihilations in the Milky Way using all-flavour neutrino-induced particle cascades. 
The analysis focuses on the DeepCore sub-detector of IceCube, and uses the surrounding IceCube strings as a veto region in order to select starting 
events in the DeepCore volume. We use 329 live-days of data from IceCube operating in its 86\hyp string configuration during 2011-2012. No neutrino 
excess is found, the final result being compatible with the background\hyp only hypothesis. From this null result, we derive upper limits on the 
velocity\hyp averaged self\hyp annihilation cross\hyp section, $\left < \sigma_A \mathrm{v} \right >$, for dark matter candidate masses ranging from 
30 GeV up to 10 TeV, assuming both a cuspy and a flat\hyp cored dark matter halo profile. For dark matter masses between 200 GeV and 10 TeV, the 
results improve on all previous IceCube results on $\left < \sigma_A \mathrm{v} \right >$, reaching a level of 10$^{-23}$ cm$^3$ s$^{-1}$, depending 
on the annihilation channel assumed, for a cusped NFW profile. The analysis demonstrates that all\hyp flavour searches are competitive with muon 
channel searches despite the intrinsically worse angular resolution of cascades compared to muon tracks in IceCube.

\keywords{Dark Matter \and Neutrino \and IceCube \and Galactic Halo \and Milky Way} 
\PACS{95.35.+d \and 12.60.-i \and 95.85.Ry \and 98.35.-a}  

\end{abstract}

\section{Introduction}
 There is strong evidence for extended halos of dark matter surrounding the visible component of galaxies. Independent indications of the existence 
of dark matter arise from gravitational effects at both galactic and galaxy\hyp cluster scales, 
as well as from the growth of primordial density fluctuations which have left their imprint on the cosmic microwave background~\cite{Lukovic:2014vma}. 
The nature of the dark matter is, however, still unknown. The most common assumption is that dark matter is composed of stable relic particles, 
whose present-day density is determined by freeze-out from thermal equilibrium as the universe expands and cools~\cite{Jungman:1995df,Feng:2010gw,Bergstrom:2012fi}. 
We focus here on a frequently considered candidate -- a cosmologically stable massive particle having only weak interactions with baryonic matter, 
namely a WIMP (Weakly Interacting Massive Particle). \par
  Within this particle dark matter paradigm, the Milky Way is expected to be embedded in a halo of WIMPs, which can annihilate  
and produce a flux of neutrinos detectable at Earth. The differential flux depends on the annihilation cross section of the WIMPs as 
\begin{equation}
\frac{\mathrm{d}\phi_{\nu}}{\mathrm{d}E} = \frac{\langle\sigma_{\mathrm{A}} \mathrm{v} \rangle}{2} \; \frac{1}{4\pi\: m^2_\chi} \; J_\mathrm{a}(\psi) \; \frac{\mathrm{d}N_\nu}{\mathrm{d}E} ,
\label{eq:NeutrinoFlux}
\end{equation}
where $\langle \sigma_\mathrm{A} \mathrm{v} \rangle $ is the product of the self-annihilation cross section, $\sigma_\mathrm{A}$, and the WIMP 
velocity, $\mathrm{v}$, averaged over the velocity distribution of WIMPS in the halo, which we assume to be spherical, $m_\chi$ is the WIMP mass, 
$\mathrm{d}N_\nu/\mathrm{d}E$ is the neutrino energy spectrum per annihilation and $J_\mathrm{a}(\psi)$ is the integral of the squared of the dark matter density along the 
line of sight. Therefore, searches for the dark matter annihilation signal in the Galactic halo can probe the WIMP self-annihilation cross-section, given their 
spatial distribution. The expected signal is particularly sensitive to the adopted density profile of the dark matter halo, which 
determines the term $J_\mathrm{a}(\psi)$ in equation~(\ref{eq:NeutrinoFlux}), $\psi$ being the angle between the direction to the Galactic Centre 
and the direction of observation~\cite{Bergstrom:1997fj,Yuksel:2007ac}. The density profile of dark matter halos determined by numerical 
simulations of structure formation is still under debate~\cite{Kravtsov:1997dp,Burkert:1995yz,Navarro:1995iw,Diemand:2009bm,Ruffini:2014zfa,deBlok:2009sp}. 
To ex\-pli\-cit\-ly quantify the effect of the choice of the halo profile on the results of our analysis, we adopt two commonly used models: 
the Navarro-Frenk-White (NFW) cusped profile~\cite{Navarro:1995iw}, and the Burkert cored profile~\cite{Burkert:1995yz,Salucci:2000ps}. We use the 
values for the parameters that characterize each profile from the Milky Way model presented in~\cite{Nesti:2013uwa}. 
The difference between the two profiles is relevant only within the Solar circle, i.e., at radii less than ~10 kpc. \par
In this paper we use data from the IceCube neutrino telescope to search for high energy neutrinos from the Galactic Centre and halo that may 
originate from dark matter annihilations. There have been several studies triggered by the observation of a electron and positron excess in the cosmic ray 
spectrum~\cite{Adriani:2008zr,Abdo:2009zk,Chang:2008aa} which favour models in which WIMPs annihilate preferably to 
leptons~\cite{Cirelli:2008pk,Donato:2008jk,Barger:2008su,Bergstrom:2009fa,Cholis:2008hb,Mandal:2009yk}. We keep, though, the analysis agnostic in terms of 
the underlying specific particle physics model that could give rise to WIMP dark matter. In this sense it is a generic approach, and our results can be 
interpreted within any model that predicts a WIMP. \par
We use data collected in 329.1 live\hyp days of detector operation between May 2011 and March 2012. 
The analysis focuses on identifying particle cascades produced by neutral or char\-ged current neutrino interactions occurring inside the 
DeepCore sub-array of IceCube, being thus sensitive to all flavours. The analysis does not explicitly try to remove muon tracks from char\-ged 
current $\nu_{\mu}$ interactions, but the event selection  has been optimized to identify and select the more spherical light pattern produced 
in the detector by particle showers. 

\section{The IceCube neutrino observatory}
\begin{figure}[!t]
\centering\includegraphics[width=0.85\linewidth,height=0.80\linewidth]{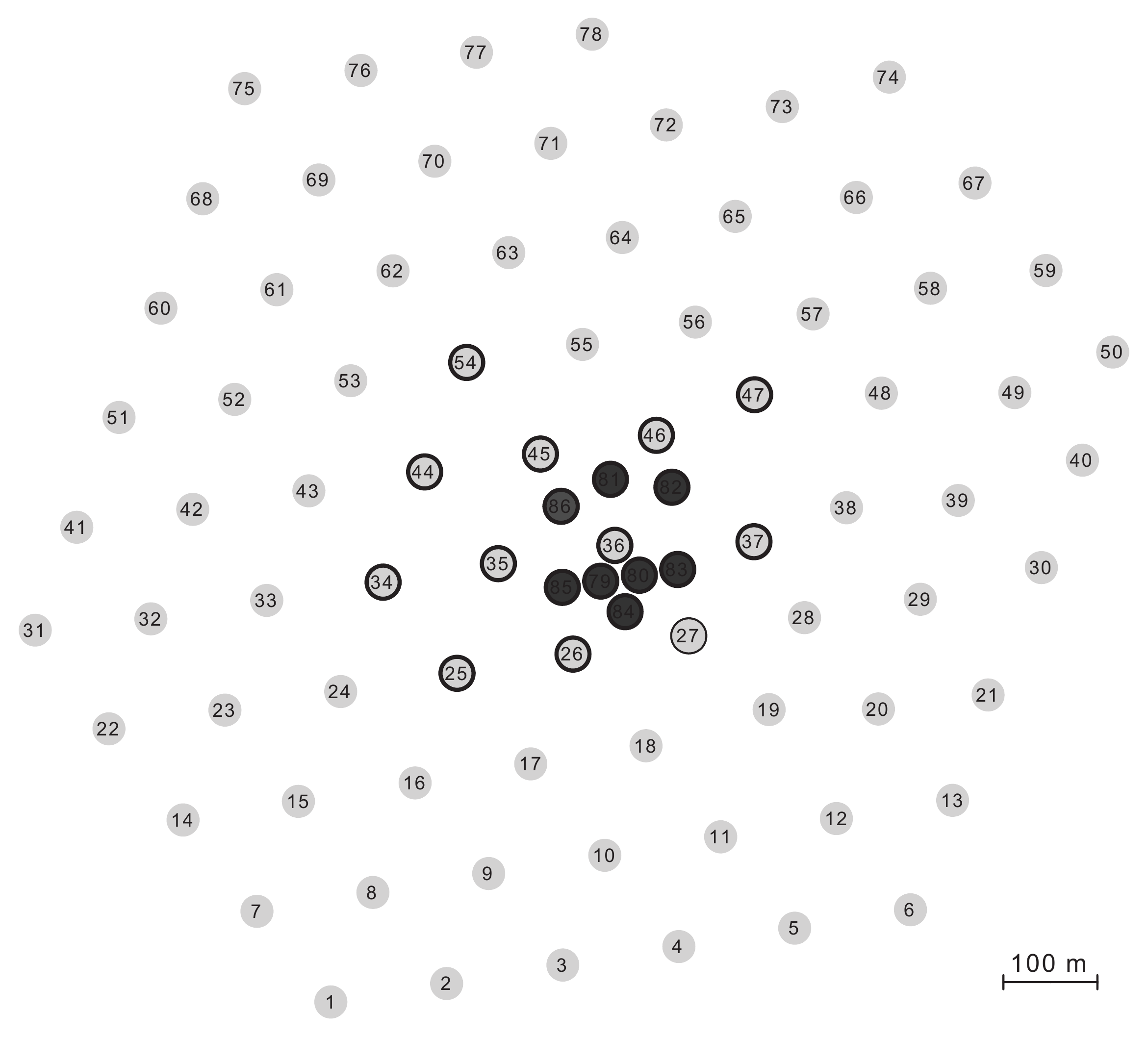}
\caption{Schematic overview of the IceCube string layout seen from above. Gray-filled markers indicate IceCube strings and black markers indicate 
the DeepCore strings with denser DOM spacing. All IceCube strings marked with a black border are included in the definition of the extended DeepCore volume used in the analysis.}
\label{fig:DC} 
\end{figure}

The IceCube Neutrino Observatory~\cite{Halzen:2010yj} is a neutrino telescope located about one kilometer from the geographical South Pole 
and consisting of an in\hyp ice array and a surface air shower array, IceTop~\cite{IceCube:2012nn}. The in\hyp ice array utilizes one cubic 
kilometer of deep ultra\hyp clear glacial ice as its detector medium. This volume is instrumented with 5160 Digital Optical Modules (DOMs) 
that register the Cherenkov photons emitted by the particles produced in neutrino interactions in the ice. The DOMs are distributed on 86 strings 
and are deployed between 1.5~km and 2.5~km below the surface. Out of the 86 strings, 78 are placed in a triangular grid of 125~m side, evenly spaced 
over the volume, and are referred to as IceCube strings. The remaining 8 strings are referred to as DeepCore strings. They are placed in between 
the central IceCube strings with a typical inter\hyp string separation of 55~m. They have a denser DOM spacing and photomultiplier tubes 
with higher quantum efficiency. These strings, along with some of the surrounding IceCube strings, form the DeepCore low\hyp energy sub-array~\cite{Collaboration:2011ym}. 
In the analysis described below, an extended definition of DeepCore was used, which includes one more layer of the surrounding IceCube strings, 
leaving a 3-string wide veto volume surrounding the fiducial volume used, see figure~\ref{fig:DC}. 
While the original IceCube array has a neutrino energy threshold of about 100~GeV, the addition of the denser infill lowers the energy threshold 
to about 10~GeV.  \par
The analysis presented in this paper uses a specific DeepCore trigger, which requires that at least three hits are registered within 2.5$\mu$s 
of each other in the nearest or next\hyp to\hyp nearest neighboring DOMs in the DeepCore sub-array. When this condition is fulfilled, the 
trigger opens a $\pm$6$\mu$s readout window centered around the trigger time, where the full in\hyp ice detector is read out. The average 
rate of this trigger is about 260~$s^{-1}$. 

\section{Signal and background simulations}\label{sec:signal}
 In order to keep the analysis general we will assume that WIMPs annihilate with 100\% branching ratio into a few benchmark channels 
($b\bar{b}$, $W^+W^-$, $\nu\bar{\nu}$, $\mu^+\mu^-$ and $\tau^+\tau^-$) and present results for these cases. Those channels effectively 
bracket the final particle spectra of realistic models with several final states. The neutrino spectra were calculated using
 PYTHIA \cite{Sjostrand:2007gs} by producing a resonance at twice the mass under consideration and forcing it to decay to the desired channel. 
The program then takes care of the further hadronization and/or decays in the standard way. We ignore the possible WIMP spin in this approach, 
which can effect the final neutrino spectrum, mainly when considering annihilations through the W$^{+}$W$^{-}$ channel~\cite{Barger:2007xf}. 
We assume that the detected neutrinos have undergone full flavour mixing given the very long oscillation baseline from the source, so there are 
equal numbers of the three flavours. The expected angular distribution of signal events in the sky is obtained by reweighting the originally 
simulated isotropic distribution by $J_\mathrm{a}(\psi)$. \par
There are two backgrounds to any search for neutrinos from the Galaxy: atmospheric neutrinos and atmospheric muons, both produced in 
cosmic\hyp ray interactions in 
the atmosphere. To estimate the effect of these backgrounds on the analysis, a sample of atmospheric muons was generated with 
the~\texttt{COR\-SI\-KA} package~\cite{Heck:1998vt} and a sample of atmospheric neutrinos was simulated with \texttt{GENIE}~\cite{Andreopoulos:2009rq} 
between 10~GeV and 200~GeV, and with \texttt{NUGEN}~\cite{Gazizov:2004va} from 200~GeV up to $10^9$~GeV, adopting the spectrum in~\cite{Honda:2006qj}. 
However, the analysis does not use background simulations to define the cuts, but instead relies on a\-zimuth\hyp scrambled data. 
This reduces the systematic uncertainties and automatically accounts for any unsimulated detector behavior. The background simulations were 
used to verify the overall validity of the analysis and the performance of the different cut levels. Since the majority of triggers in IceCube 
are due to atmospheric muons, the distributions of the variables used in the analysis must agree between data and the \texttt{CORSIKA} simulation 
at early selection levels, while at higher selection levels the data should show a significant fraction of atmospheric neutrinos.
 Atmospheric muons and particles resulting from neutrino interactions in or near the detector are propagated through the detector volume and 
their Cherenkov light emission simulated. Cherenkov photons are then propagated through the ice using the \texttt{PPC} package~\cite{Chirkin:2013tma}, 
and the response of the detector calculated. From this point onwards, simulations and data are treated identically through further filtering and data cleaning.
\begin{figure*}[!t]
\includegraphics[width=0.45\linewidth,height=0.35\linewidth]{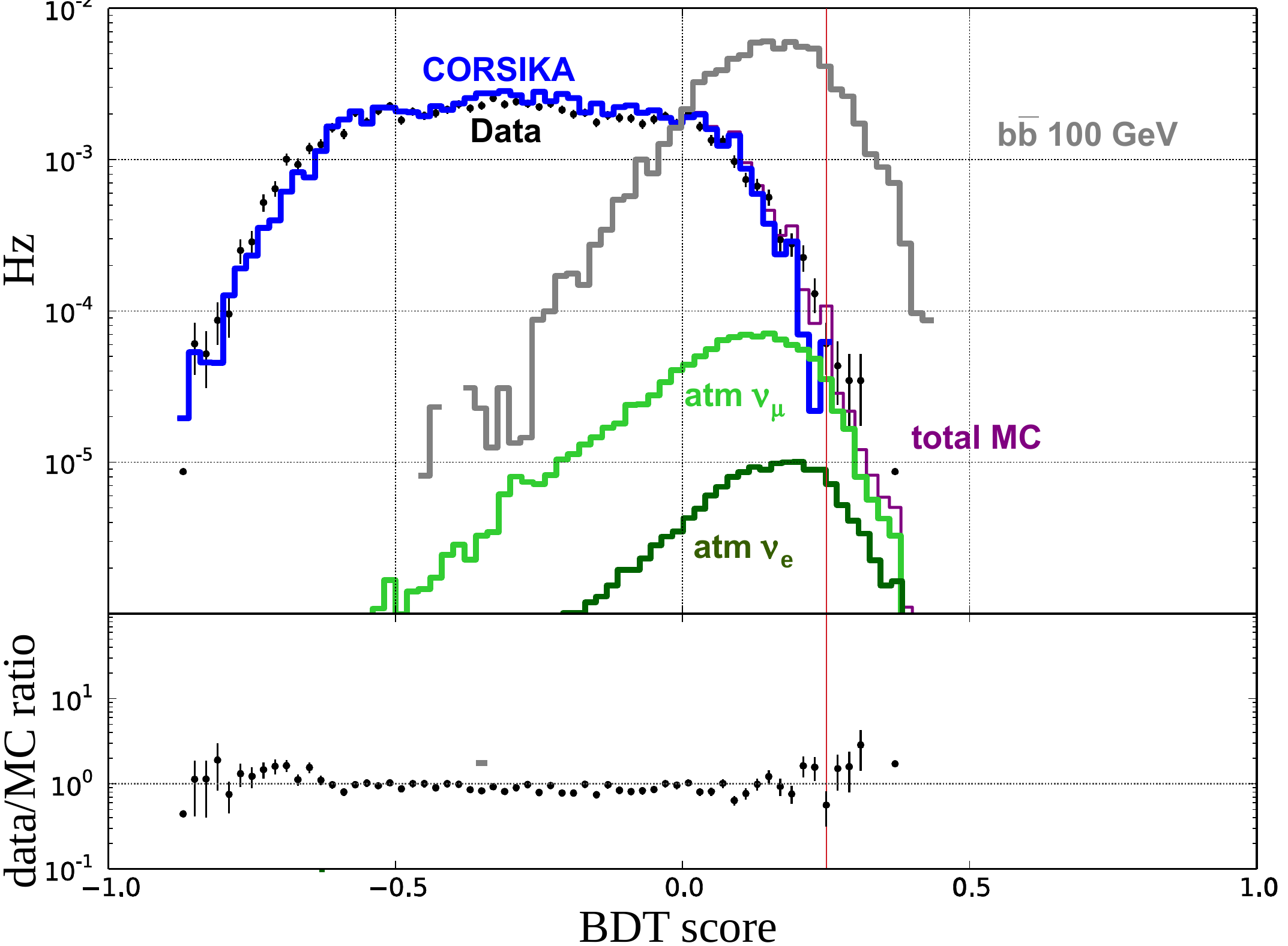}
\hfill
\includegraphics[width=0.45\linewidth,height=0.35\linewidth]{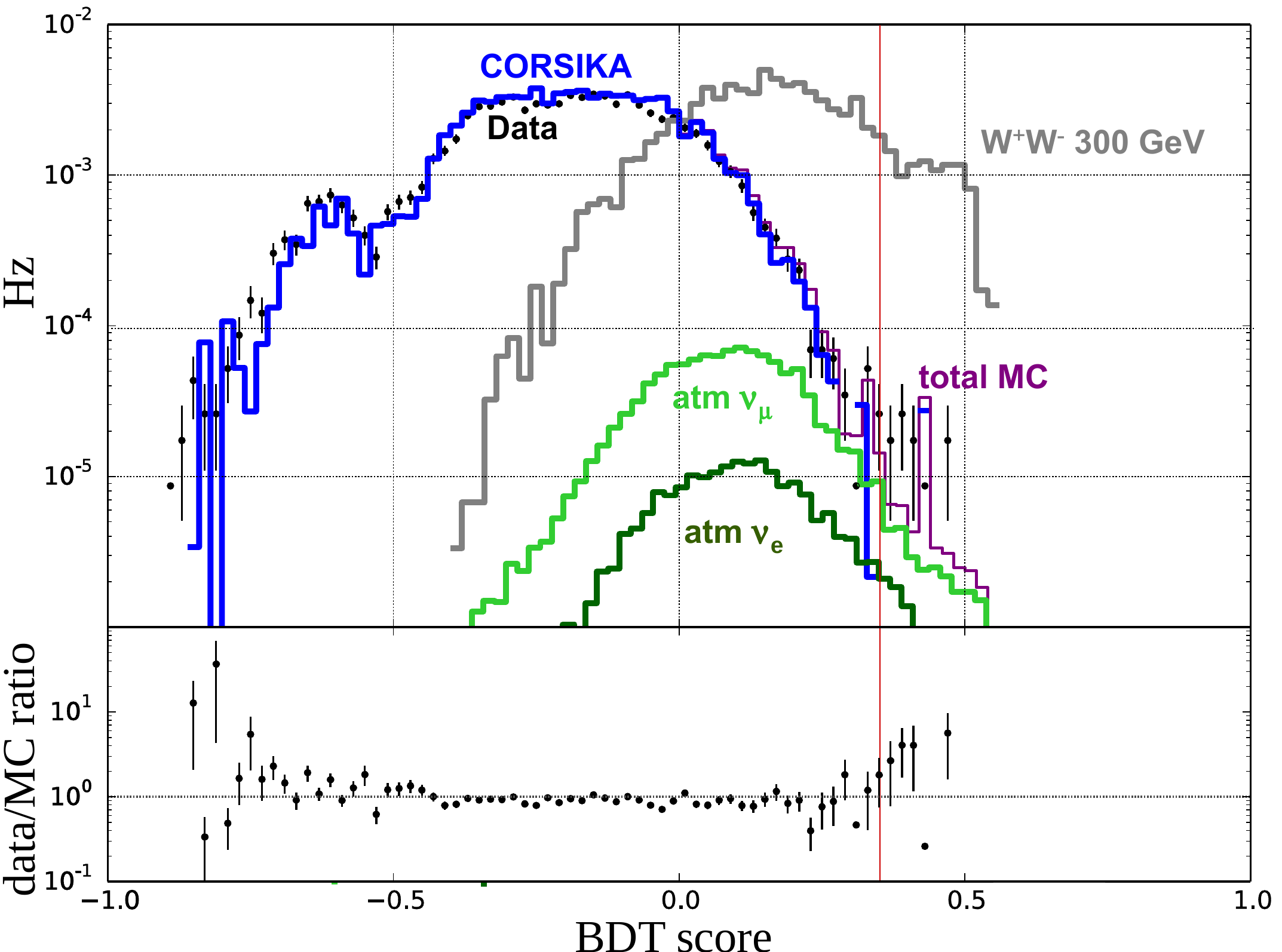}
\caption{The score distribution for the BDT trained on the $b\bar{b}$ 100~GeV signal channel (left) and 
for the BDT trained on the $W^+W^-$ 300~GeV signal channel (right). The plot shows the passing event rate (in Hz) for simulated atmospheric 
muons (blue line) and atmospheric neutrinos (green lines), as well as for the sum of these two components (total MC, purple line), 
compared with the data passing rate. The passing rate for each signal channel the BDT was trained for is also shown (grey lines), normalized to the 
experimental data rate. The final cuts on the score are marked with vertical lines. Events were kept if any of the scores were above the cut values.   
The lower panel in each plot shows the ratio of the data passing rate to the total expected background.}
\label{fig:bdts} 
\end{figure*}

\section{Data selection}\label{sec:selection}
 The triggered data are first cleaned of potential noise hits that could effect the performance of the track and cascade reconstruction 
algorithms. Hits that lie outside a predetermined time window around the trigger time or which do not have another causally connected hit within 
a predefined radius, are removed. The data is then filtered by a fast algorithm that selects events starting in the DeepCore fiducial volume, 
in order to remove events triggered by through--going atmospheric muons. The IceCube strings surrounding DeepCore are used as a veto for 
incoming tracks. The algorithm selects events with the ``amplitude\hyp weighted'' centre of gravity of all hits inside the DeepCore 
volume~\footnote{The amplitude\hyp weighted centre of gravity of an event is defined as $\vec{r}_{\texttt{COG}}=\sum a_i \vec{r}_i / \sum a_i$, 
where $a_i$ and $\vec{r}_i$ are the amplitude and position of the $i$th hit. The sum runs over all the hits in the event (after hit cleaning).}, 
and no more than one hit in the surrounding IceCube strings causally connected 
with that point. This filter reduces the passing event rate by nearly a factor of 10.\par
The event sample is further reduced by requiring a minimum number of eight hits in the event distributed in at least four strings. This 
ensures that the remaining events can be well reconstructed. The events are then processed through a series of reconstructions aimed  at 
determining their type (cascade or track), arrival direction and energy. In a first stage, two first\hyp guess reconstructions are applied; 
fits for a track hypothesis and for a cascade hypothesis are performed in order to obtain a quick characterization of the events and perform 
a first event selection. These fits are based on the position and time of the hits in the detector, but do not include information about 
the optical properties of the ice, in order to speed up the computation. The track hypothesis performs a $\chi^2$ fit of a straight line 
to the hit pattern of the event, returning a vertex and a velocity~\cite{Aartsen:2013bfa}. The cascade hypothesis is based on determining 
the amplitude\hyp weighted centre of gravity of the hits in the event and its associated time. The algorithm calculates the three principal 
axes of the ellipsoid spanned by the spacial distribution of hits, and the longest principal axis is selected to determine the generic 
direction of the event. Since the specific incoming direction along the selected axis is ambiguous, the hit times are projected onto this 
axis, from latest to earliest, to characterize the time-development of the track so that it points towards where the incident particle originated. 
The tensor of inertia reconstruction is generally only suitable as a first guess of the direction for track\hyp like events, since for 
cascade\hyp like events the three principal axes of the ellipsoid will be close to equal in size. This property, however, can be used to 
discriminate between tracks and cascades.  Additionally, a series of cuts based on variables derived from the geometrical distribution of hits, 
as well as from information from the first guess reconstructions, are applied. These cuts bring the experimental data rate down by a factor of 
about 3000 with respect to trigger level, while keeping about 50\% of the signal, depending on the WIMP mass and annihilation channel considered.

\begin{figure*}[t]
\centering\includegraphics[width=0.45\linewidth,height=0.35\linewidth]{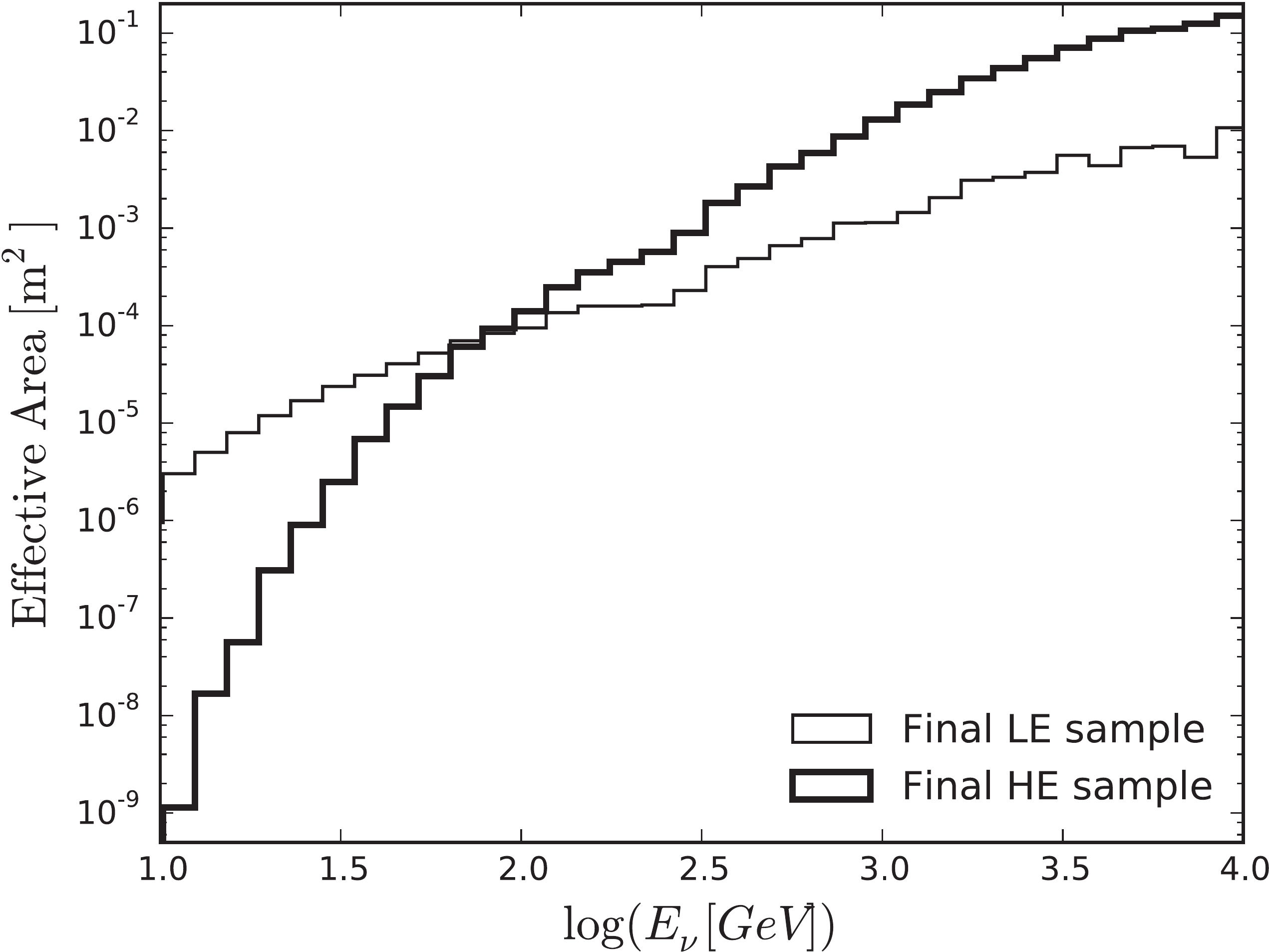}
\hfill
\centering\includegraphics[width=0.45\linewidth,height=0.35\linewidth]{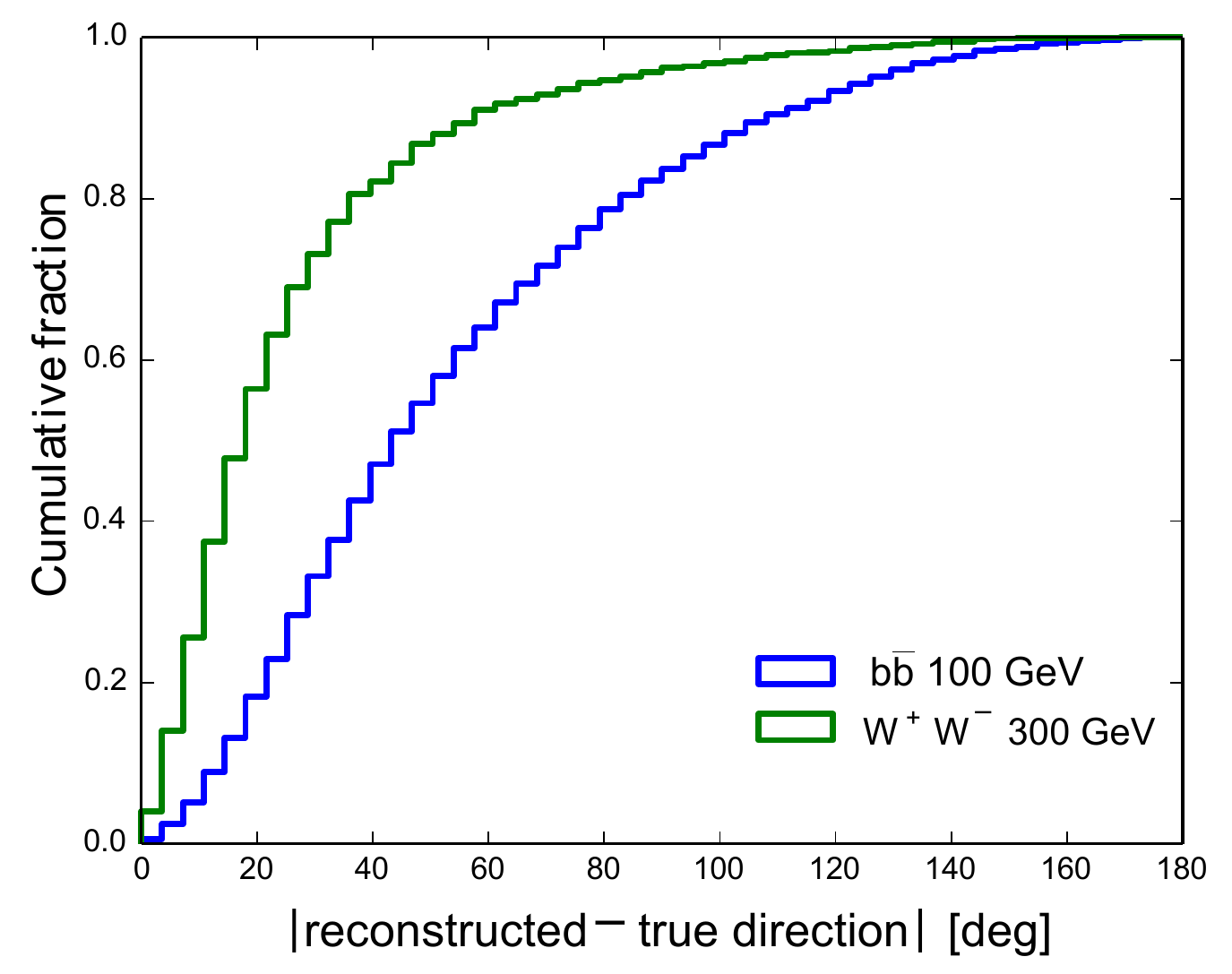}
\caption{{\bf Left:} All\hyp flavour neutrino effective area as a function of energy for the two event selections of the analysis, the low\hyp energy (LE) and 
high\hyp energy (HE) selections.
{\bf Right:} Cumulative angular resolution (based on the space angle between the reconstructed and true direction of incoming neutrinos) at final analysis level.}
\label{fig:aeff_angres} 
\end{figure*}

\begin{figure*}[!t]
\includegraphics[width=0.45\linewidth,height=0.35\linewidth]{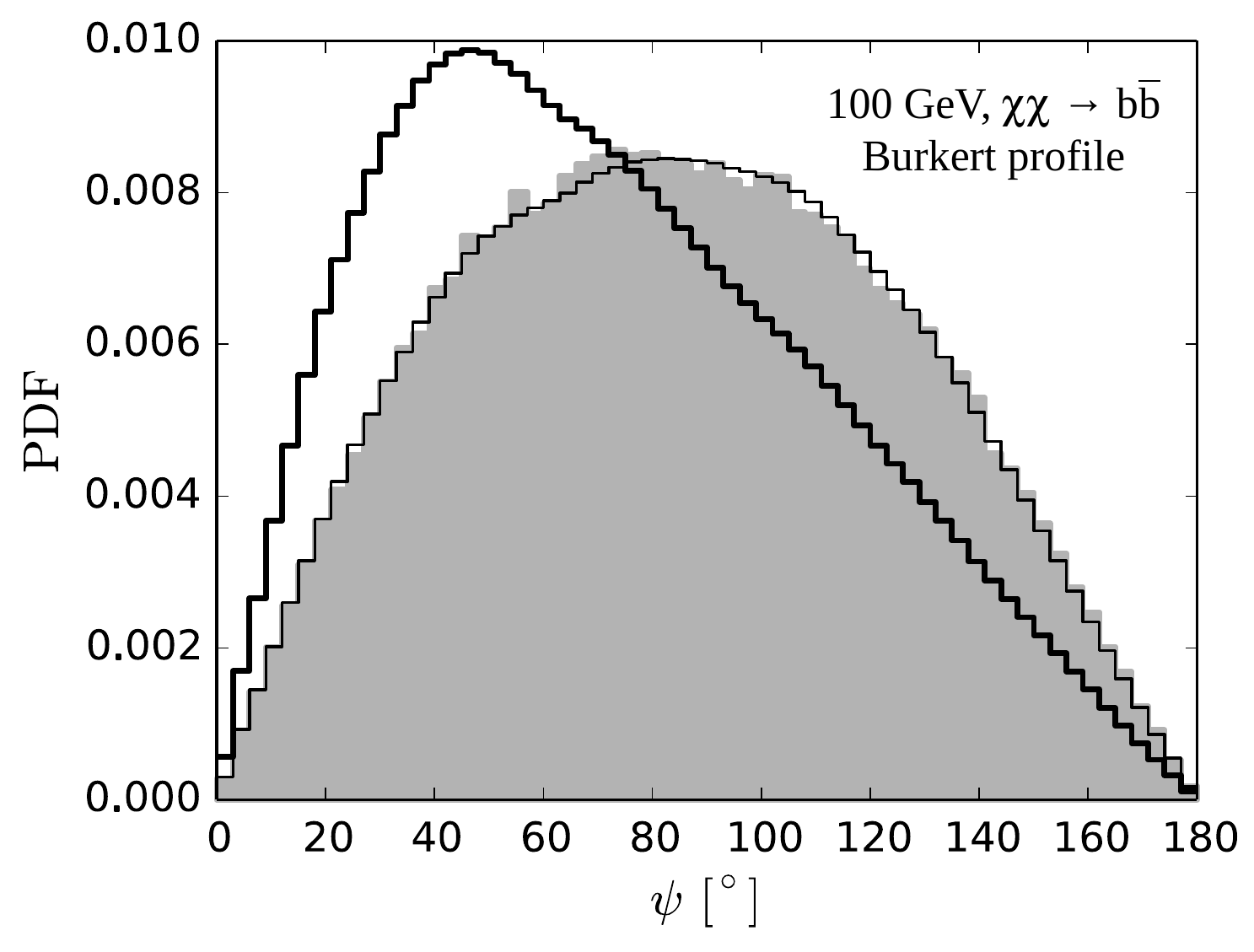}
\hfill
\includegraphics[width=0.45\linewidth,height=0.35\linewidth]{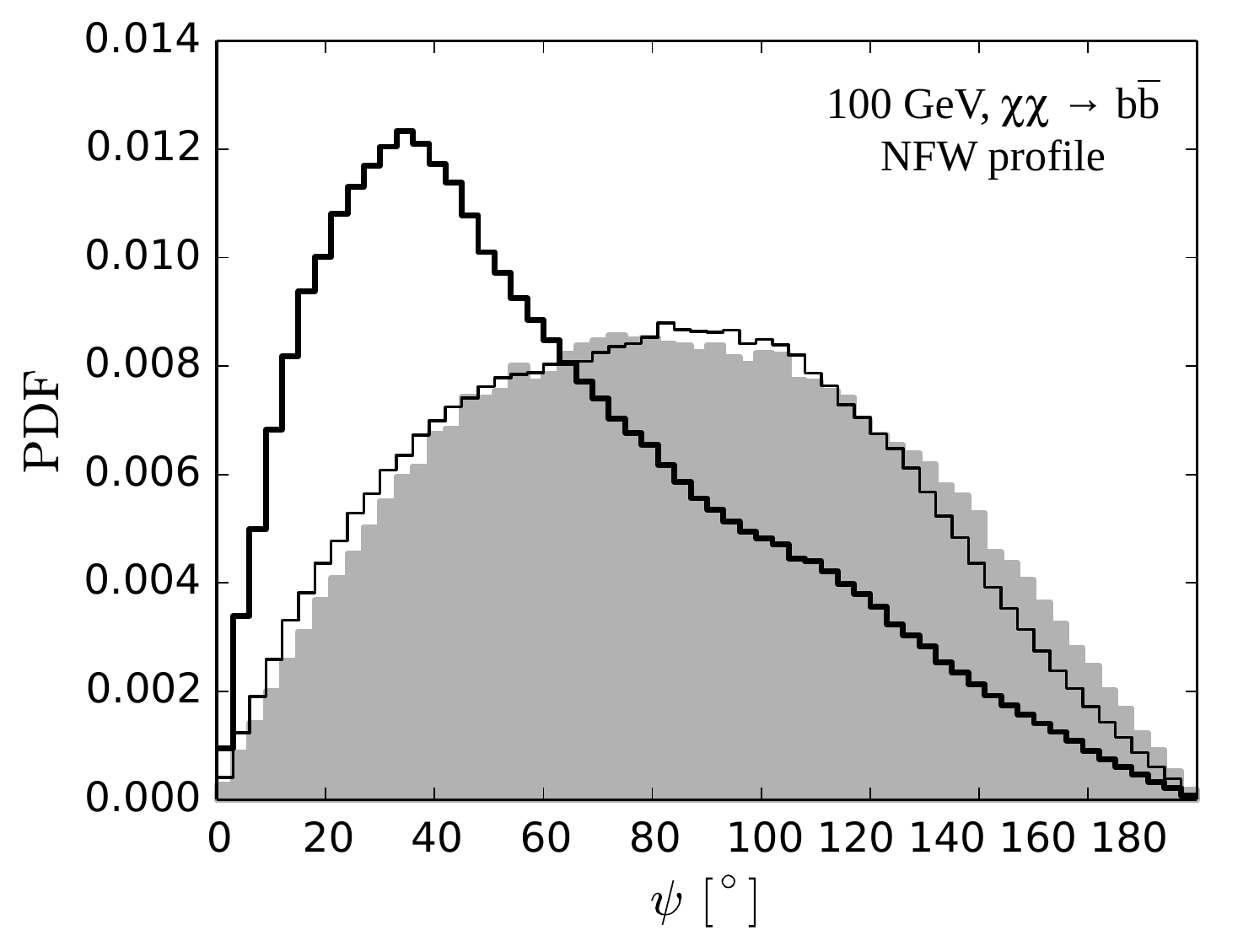}\\
\caption{Example of the space angle PDF for one of the signal channels considered ($\chi \chi \rightarrow b\bar{b}$) and two halo profiles, the Burkert profile (left) 
and the NFW profile (right). In each plot the signal PDF, $f_\mathrm{S}(\psi)$, is shown as a 
thick black line, and the two components of the background PDF, the scrambled data, $f_\mathrm{sd}(\psi)$,  and the scrambled signal, $f_\mathrm{ss}(\psi)$,
are shown as the gray shaded area and the thin black line, respectively. The angle $\psi$ represents the angular distance between the direction of reconstructed tracks  
and the location of the Galactic Center.
\label{fig:allPDFs}}
\end{figure*}

\begin{table*}
\caption{Data rates at different cut levels. For the two benchmark signal channels, 
100~GeV $b\bar{b}$ and  300~GeV $W^+W^-$, values are given as percentage of signal retention relative to the DeepCore and event--quality filter level. The livetime of the experimental data set is 329.1 days.}
\label{tab:rates}
\begin{tabular*}{\textwidth}{@{\extracolsep{\fill}}lrrrrrrr@{}}
\hline
 & \multicolumn{1}{c}{Exp. data} & \multicolumn{1}{c}{Atm $\mu$} & \multicolumn{1}{c}{Atm. $\nu_e$} & \multicolumn{1}{c}{Atm. $\nu_{\mu}$} & \multicolumn{1}{c}{Total atm. $\nu$} & \multicolumn{1}{c}{100~GeV $b\bar{b}$} & \multicolumn{1}{c}{300~GeV$W^+W^-$} \\
 & \multicolumn{1}{c}{($s^{-1}$) } & \multicolumn{1}{c}{ ($s^{-1}$)} & \multicolumn{1}{c}{ ($s^{-1}$)} & \multicolumn{1}{c}{ ($s^{-1}$)} & \multicolumn{1}{c}{($s^{-1}$)} & \multicolumn{1}{c}{} & \multicolumn{1}{c}{} \\
 
\hline
Trigger & $\sim$260 &  & &  &   &    &    \\
DeepCore \& event--quality filter  &   $\sim$18   & $\sim$17 &  &  &   & 100\%  & 100\%\\
pre-BDT linear cuts &   8.07$\times$10$^{-2}$ & 8.89$\times$10$^{-2}$ &  2.11$\times$10$^{-4}$ & 1.12$\times$10$^{-3}$ & 1.33$\times$10$^{-3}$ & 51.0\% & 46.0\% \\
BDT$_\text{LE}$        &  2.06$\times$10$^{-4}$ & 4$\times$10$^{-5}$ & 2.58$\times$10$^{-5}$ & 7.74$\times$10$^{-5}$ & 1.03$\times$10$^{-4}$  & 7.78\% & 2.85\%\\
BDT$_\text{HE}$       &  7.61$\times$10$^{-5}$ & 2$\times$10$^{-5}$ & 1.02$\times$10$^{-5}$ & 2.56$\times$10$^{-5}$ & 3.58$\times$10$^{-5}$ &  0.77\%  & 5.84\%\\
\hline
\end{tabular*}
\end{table*}

\begin{table*}[t]
\caption{Summary of systematic uncertainties for both the low\hyp energy (LE) and high\hyp energy (HE) event selections presented for both halo profiles used in the 
analysis. The total is the quadratic sum of each individual contribution}
\label{tab:systematics}
\begin{tabular*}{\textwidth}{@{\extracolsep{\fill}}lrrrr@{}}
\hline
  & \multicolumn{2}{c}{Burkert profile} & \multicolumn{2}{c}{NFW profile} \\
 & \multicolumn{1}{l}{LE selection} & \multicolumn{1}{l}{HE selection} & \multicolumn{1}{l}{LE selection} & \multicolumn{1}{l}{HE selection}  \\
 
\hline
Ice optical properties & 8\% & 8\% & 12\%  & 12\%  \\
Hole ice optical properties &  24\%   & 15\%  & 24\%  & 10\%  \\
       &  &  &  &  \\
DOM efficiency    & 17\%  & 10\%  & 35\%  & 12\%    \\
Noise model       &  10\% & 5\% & 8\%  & 10\%    \\
Detector calibration    & $<$5\%  & $<$5\% &  $<$5\% & $<$5\%   \\
Analysis    & 2\%  & 2\% &  2\% & 2\%   \\
       &  &  &  &  \\
{\bf Total}    & {\bf 34\%}  & {\bf 21\%} &  {\bf 45\%} & {\bf 23\%}   \\
\hline
\end{tabular*}
\end{table*}
At this point three sophisticated likelihood\hyp based reconstructions are applied on all the remaining events. The likelihood reconstructions 
aim at determining a set of parameters ${\bf{a}}=({\bf{x}_0}, t_0, \theta, E_0)$ given a set of measured data points ${d_i}$ (e.g. time and 
spatial coordinates of every hit in an event). Here $\vec{x}_0$ is an arbitrary point along the track, $t_0$ is the event time at position 
${\bf{x}_0}$, $\theta$ is the direction of the incoming particle and $E_0$ is the deposited energy of the event.
 The reconstructions attempt to find the value of ${\bf{a}}$ that maximizes the likelihood function, which is based on the Probability Density 
Function (PDF) of measuring the data point ${d_i}$ given the set of parameters ${\bf{a}}$. 
For a cascade reconstruction there are seven degrees of freedom, while an infinite track reconstruction has only six since the point ${\bf{x}_0}$ can be
chosen arbitrarily along the track. The first reconstruction is based on an infinite track hypothesis, fitting 
only direction, not energy. The second reconstruction uses a cascade hypothesis, and it fits for the vertex position, direction and energy of 
the cascade. These two reconstructions use an analytic approximation for the expected hit times in the DOMs given a track or cascade 
hypothesis~\cite{Ahrens:2003fg}, rather than a full description of the optical properties of the ice. 
Since the focus of this analysis is to identify cascades, an additional, more advanced, cascade reconstruction is also performed, using the 
previous one as a seed. This second cascade reconstruction uses the full description of the optical properties of the Antarctic ice, as well 
as information of the position of non\hyp hit DOMs through a term added to the energy likelihood. 
The three likelihood reconstructions return the best fit values of the variables of the vector ${\bf{a}}$ they fit for, as well as a likelihood 
value of their respective hypothesis, which is used in a further selection of events using linear cuts.\par
  The final selection of events uses Boosted Decision Trees (BDT)~\cite{Hocker:2007ht} to classify events as signal\hyp like or background\hyp like. 
Two BDTs were trained using data as background and a different benchmark reference signal each. One of the BDTs (BDT$_\text{LE}$) was trained using the neutrino 
spectrum from a 100~GeV WIMP annihilating into $b\bar{b}$, while the other, BDT$_\text{HE}$, was trained on the neutrino spectrum of a 300~GeV WIMP annihilating 
into $W^+W^-$. These two spectra were chosen to represent a soft and hard neutrino spectrum 
respectively, so the sensitivity of the analysis to other WIMP masses and/or annihilation channels with similar spectra can be evaluated with the same 
cuts on the BDT output scores. This removes the need to train a different BDT specifically for each mass and annihilation channel.  Since no variables depending 
on the arrival direction of the events are used in the BDT training, the event sample is kept blind with respect to the position of the Galactic Centre in the sky. \par
  Seven variables that showed a good separation power between signal and background, selected among an initial larger set of variables that were tried, 
were used to train the BDTs. The variables are based on the different geometrical patterns that tracks and cascades leave in the detector, as well 
as on their different time development.  The whole data set was classified by the two BDTs so each event was assigned two BDT scores. In order to decide on the best cut 
value on each BDT output, the range of BDT score values was scanned and the sensitivity of the analysis was calculated for each of them. The scores 
producing the best sensitivity for each of the two signal channels for which the BDTs were trained were selected. Events with a BDT$_\text{LE}$ score above 
the optimal value are referred to as the ``low\hyp energy'' (LE) sample, and events with a BDT$_\text{HE}$ score above the corresponding cut value are 
referred to as the ``high\hyp energy'' (HE) sample.  The remaining number of events in each sample is 5892 events in the LE sample and 2178 events in the HE sample. 
The overlap between the two samples (events which have both BDT scores above the respective cut values) is 664 events. The final BDT score distributions for 
the 100~GeV $b\bar{b}$ and the 300~GeV  $W^+W^−$ channels are 
presented in figure~\ref{fig:bdts}, with the vertical lines marking the optimal cut values used to select the final event sample.

\begin{figure*}[t]
\includegraphics[width=0.45\linewidth,height=0.35\linewidth]{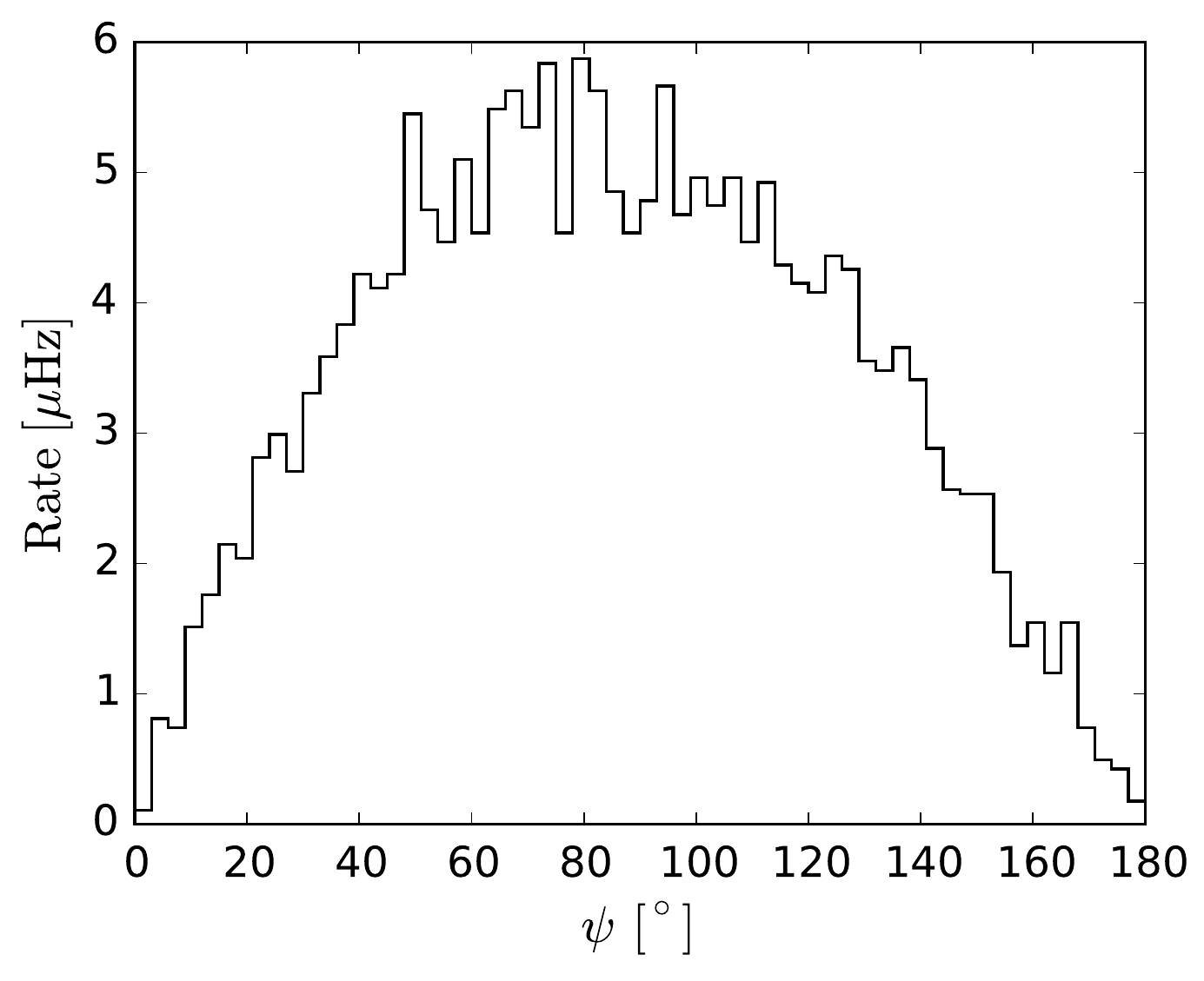}
\hfill
\includegraphics[width=0.45\linewidth,height=0.35\linewidth]{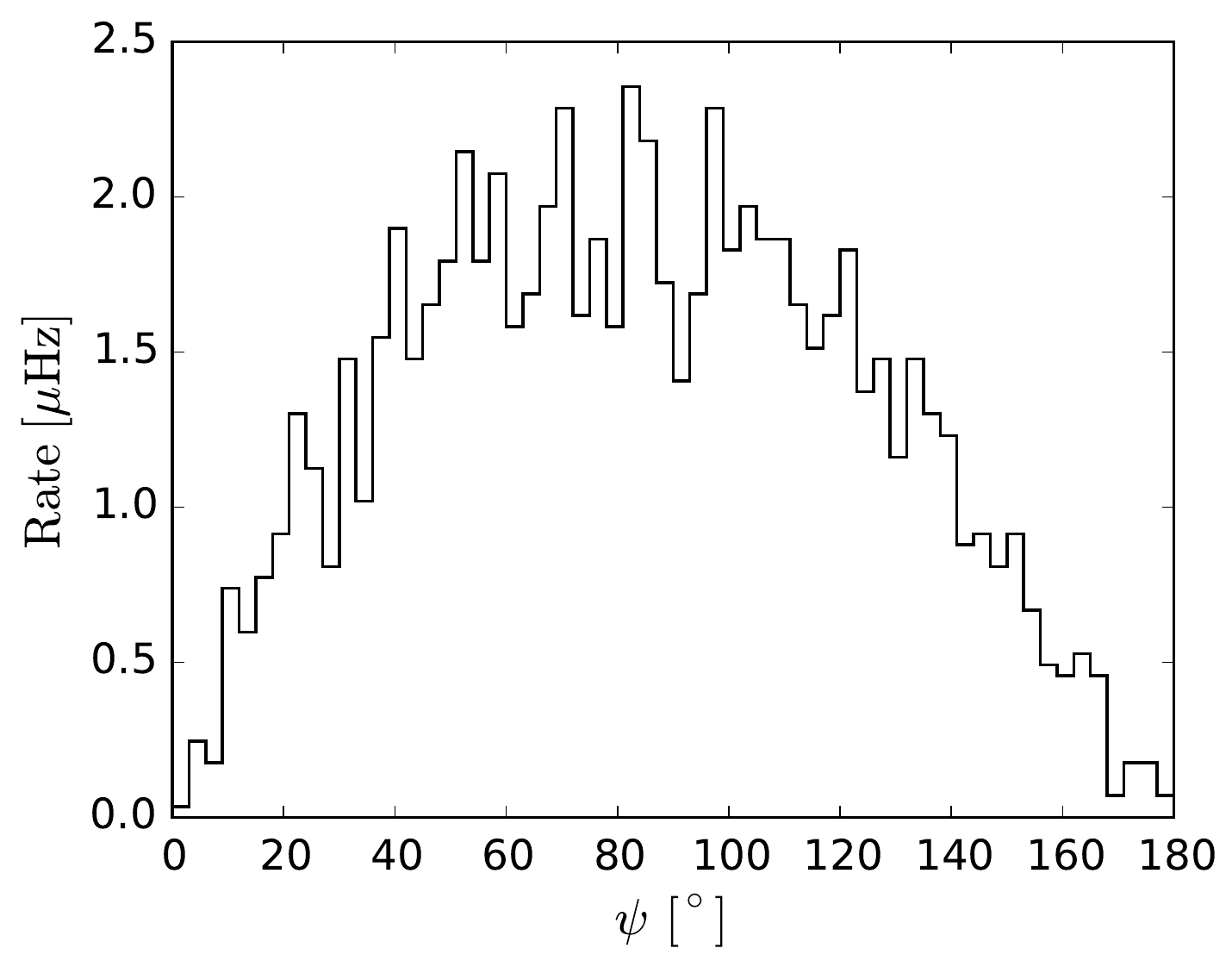}\\
\caption{Distributions of the $\psi$ angles of the final event samples. The bin contents are directly proportional to the number of observed 
events, to which we choose not to assign any statistical uncertainty. {\bf Left:} The Low Energy (LE) sample, which contains 5892 observed events. 
{\bf Right:} The High Energy (HE) sample, which contains 2178 observed events. }
\label{fig:ang_dist}
\end{figure*}

After the BDT classification, the data has been reduced by a factor of about $1(3)\times 10^{6}$ for the LE(HE) sample, but still contains about 20\% of atmospheric 
muon contamination. The remaining signal in the two benchmark scenarios considered amounts to about 6\%(8\%) respectively. A summary of 
the event selection rates, as well as signal efficiency, is given in table~\ref{tab:rates}. The effective area for the two event selections, a 
measure of how efficient the detector is for the present analysis, is shown in the left plot of figure~\ref{fig:aeff_angres}. The right plot in the same 
figure shows the cumulative angular resolution (space angle between the reconstructed and true direction of the incoming neutrino) for the two benchmark channels 
used in training the BDTs.

\section{Systematic uncertainties}\label{sec:systematics}
 In order to estimate the effect of experimental systematic uncertainties on the final sensitivity, Monte Carlo simulation studies were done, where the parameters 
defining a given input were varied within their estimated uncertainty. The main source of systematic uncertainties is the limited knowledge of the optical properties 
of the ice, both the bulk ice between 1450~m and 2500~m, as well as the ``hole ice'', i.e. the ice that forms as the water in the hole drilled for the string 
deployment refreezes. The scattering and absorption coefficients of the ice as a function of depth have been determined by in\hyp situ flash measurements, 
and a standard ``ice model'' for IceCube has been derived~\cite{Aartsen:2013rt}. The effect on the uncertainty of the estimated absorption and scattering length 
was investigated by varying the baseline settings by $\pm$~10\% individually. 
Their contribution to the uncertainty on the sensitivity lies in the range 8\%-12\%. Furthermore, there are indications that the hole ice contains residual air 
bubbles that result in a shorter scattering length in this ice compared to the ancient glacial bulk ice surrounding it.  In the baseline simulation data sets the 
scattering length of the hole ice is set to 50~cm. Varying this parameter between 30~cm and 100~cm yields a 10\%-24\% change on the sensitivity. Recently, a more 
detailed modeling of the bulk ice  has been developed~\cite{Dima:13aa}. It includes anisotropic scattering and accounts for the tilt of the different ice layers 
across the IceCube volume. Preliminary studies indicate that the effect on the sensitivity of this model is negligible for high\hyp energy events, but it 
can be sizable for the lowest\hyp energy events, reducing the sensitivity for low WIMP masses up to 25\%. These effects have not been included in this analysis.\par
  The overall efficiency of the process of converting the Cherenkov light into a detectable electrical signal by the DOM is another source of uncertainty. This 
effect was investigated by changing the DOM efficiency in the signal simulation by $\pm$10\%, according to measurements of the performance of the DOMs in 
laboratory tests before deployment, as well as in in\hyp situ calibration measurements after deployment. This uncertainty translates into an uncertainty on 
the final sensitivity of 10\% -- 35\%, depending on event selection. The effect is stronger for low\hyp energy events that can fall under the detector threshold 
if less light is being captured. Additional, but minor, effects arise from the implementation of the photomultiplier dark noise in the simulation, the timing and 
geometry calibration of the detector and from the intrinsic randomness of several steps of the analysis, like time-scrambling of the data or the many pseudo-experiments 
performed to calculate the sensitivity.  \par
  All systematic uncertainties considered are summarized in table~\ref{tab:systematics} together with the total (quadratic sum) for the low and high\hyp energy
selections for both halo profiles. In order to be conservative, the limits presented in section~\ref{sec:analysis} for each WIMP mass and annihilation 
channel were rescaled by the corresponding total systematic uncertainty shown in table~\ref{tab:systematics}. 

\begin{figure*}[t]
\includegraphics[width=0.45\linewidth,height=0.35\linewidth]{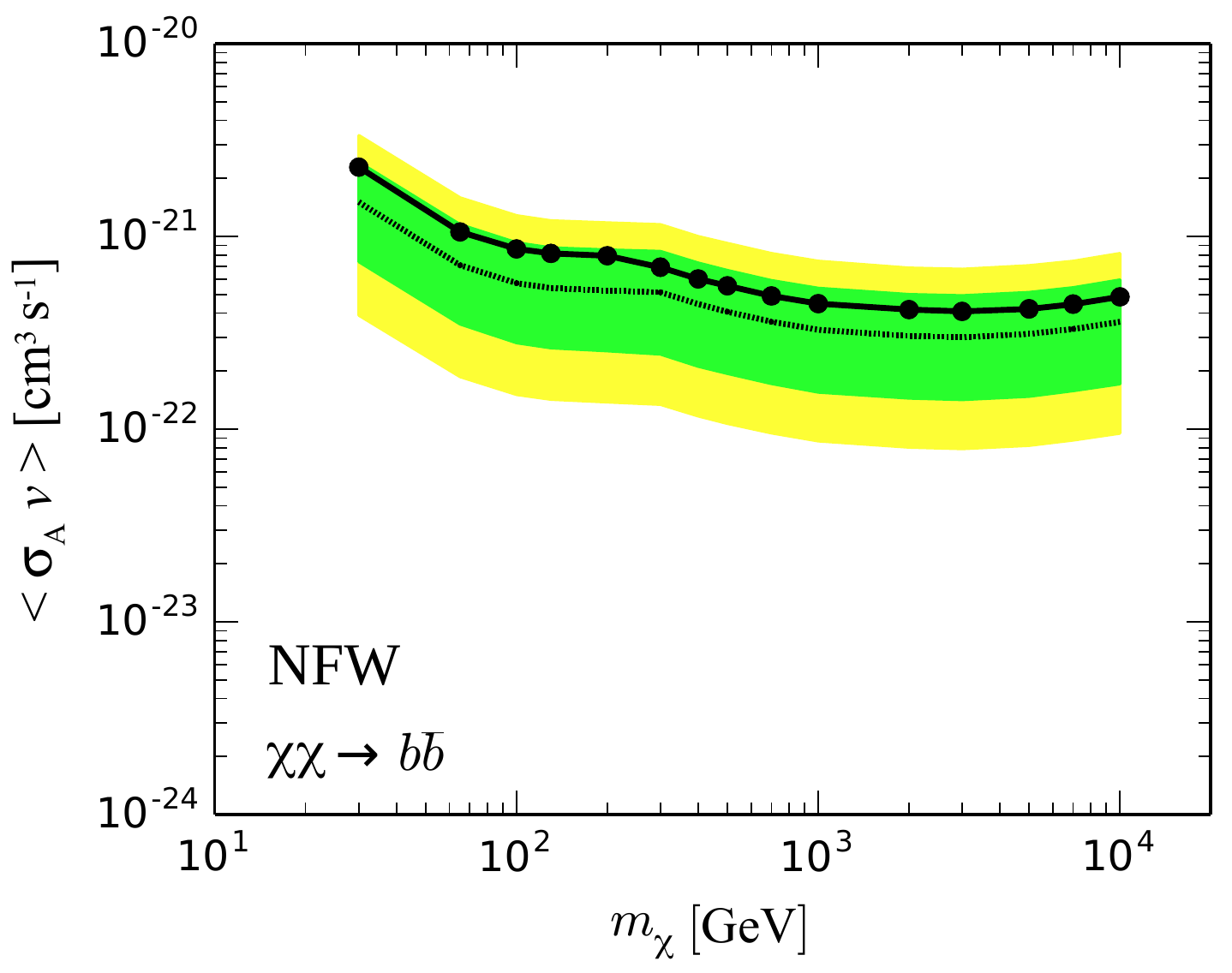}
\hfill
\includegraphics[width=0.45\linewidth,height=0.35\linewidth]{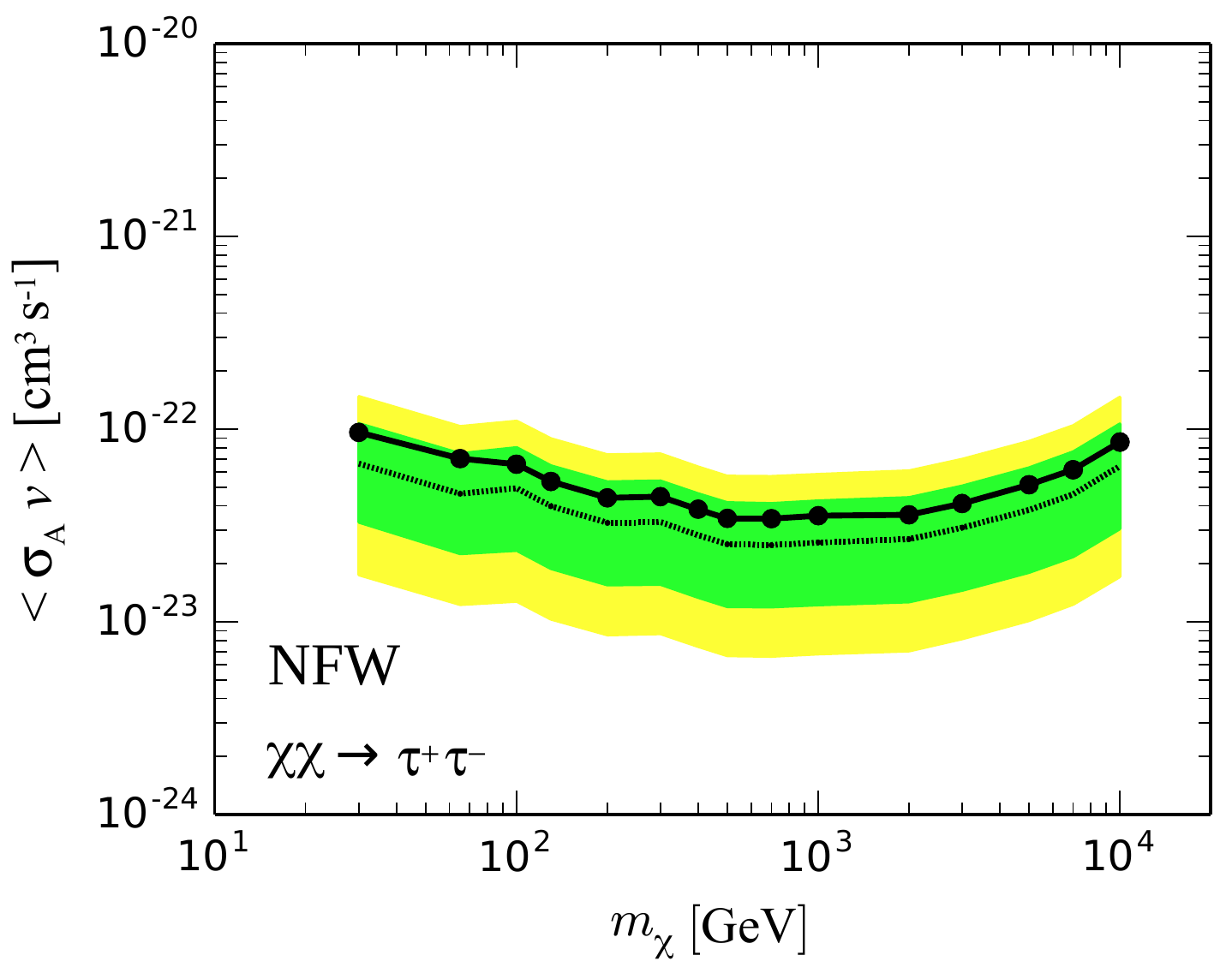}\\
\includegraphics[width=0.45\linewidth,height=0.35\linewidth]{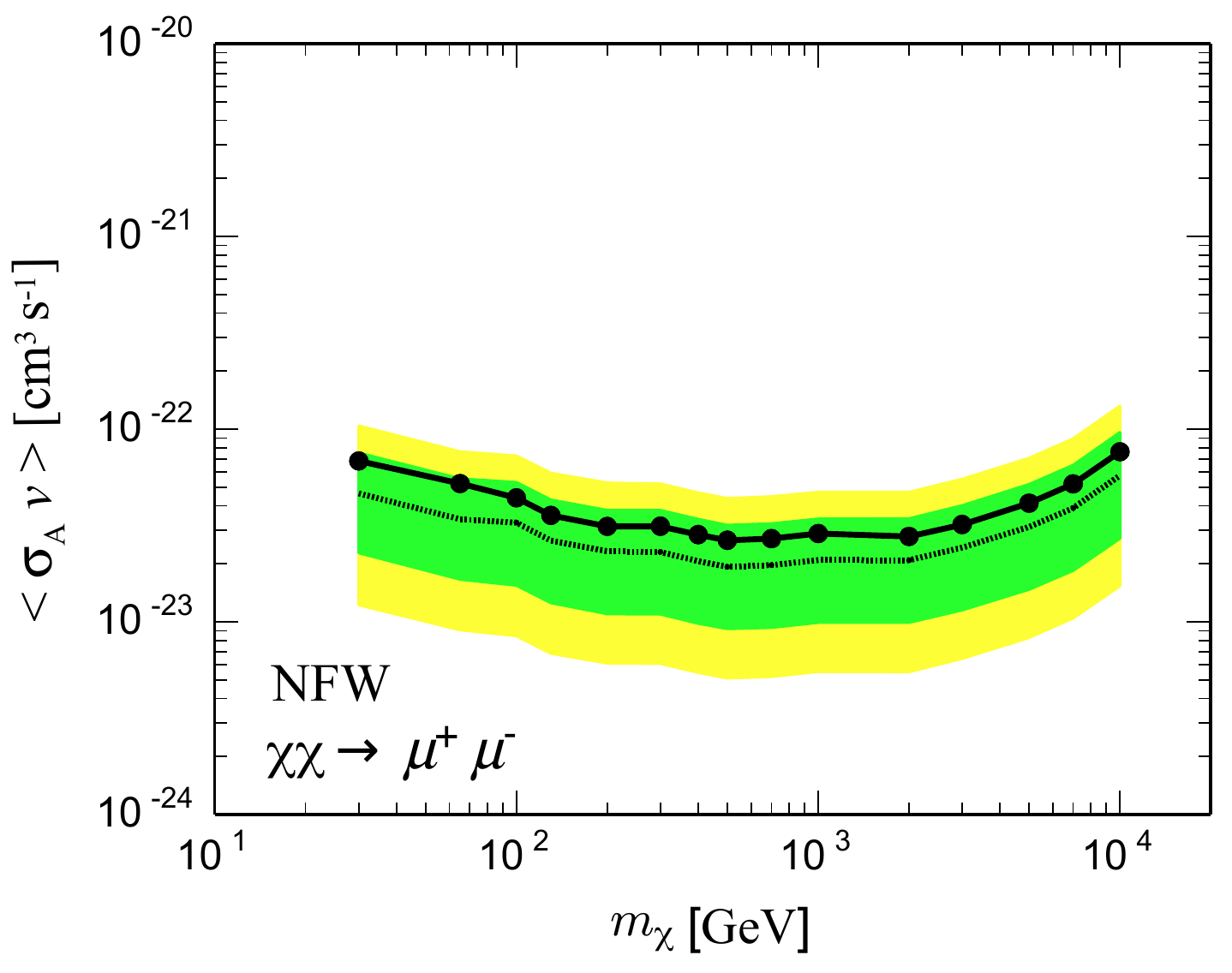}
\hfill
\includegraphics[width=0.45\linewidth,height=0.35\linewidth]{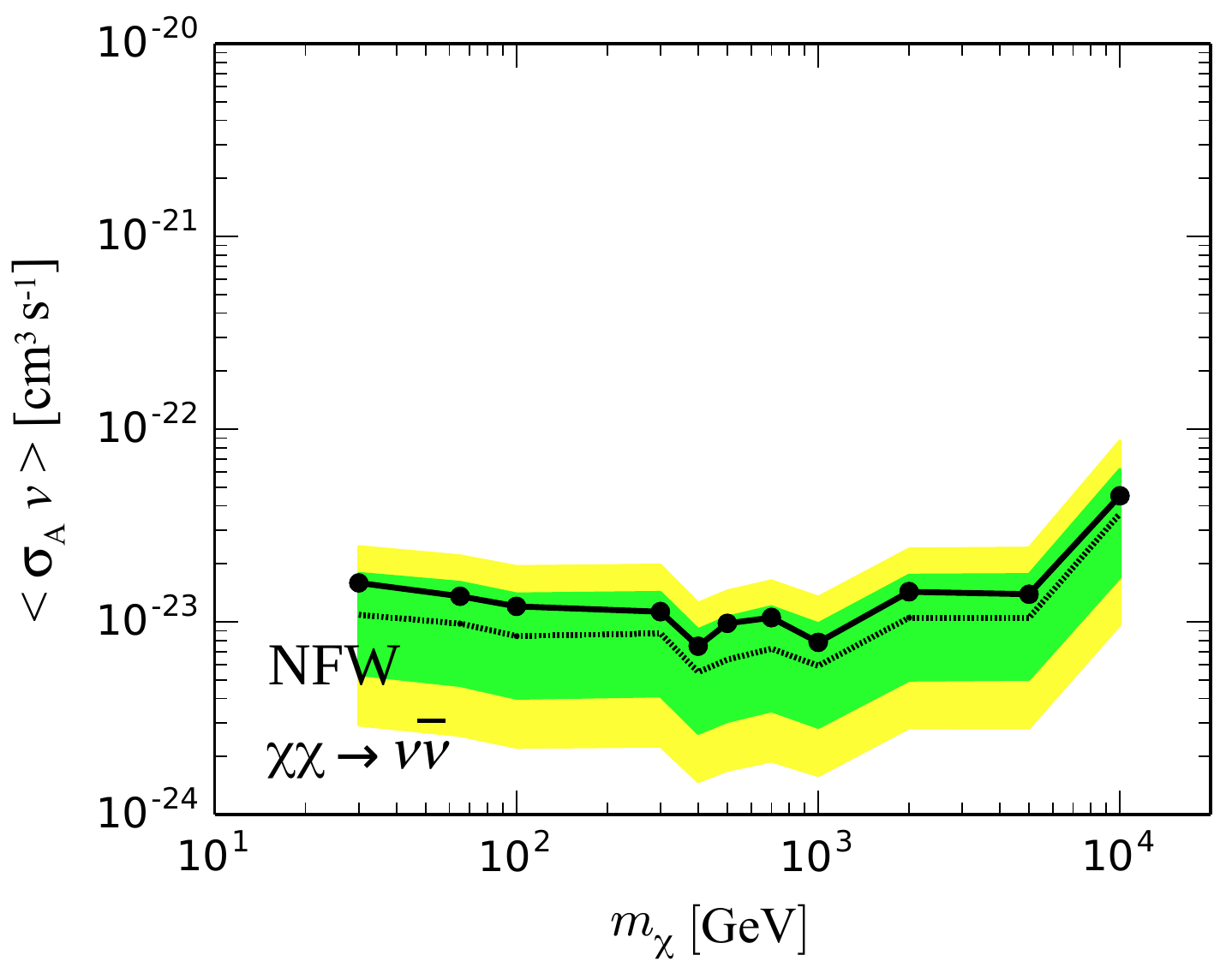}
\caption{Upper limits (90\% C.L., solid black line) on the velocity\hyp averaged WIMP self\hyp annihilation cross section, $\langle\sigma_{\mathrm{A}} \mathrm{v}\rangle$, for 
the NFW halo model together with the corresponding sensitivities (dashed black line) and their 1$\sigma$ (green) and 2$\sigma$ (yellow) statistical uncertainties. The black 
dots represent the masses probed, while the black line in between is drawn to guide the eye. Each plot corresponds to a different annihilation channel as indicated in 
the legend. The local dark matter density used was $\rho_{\mathrm{local}}$ = 0.47 GeV/cm$^3$~\cite{Nesti:2013uwa}.}
\label{fig:UpperLimitsNFW}
\end{figure*}

\begin{figure*}[t]
\includegraphics[width=0.45\linewidth,height=0.35\linewidth]{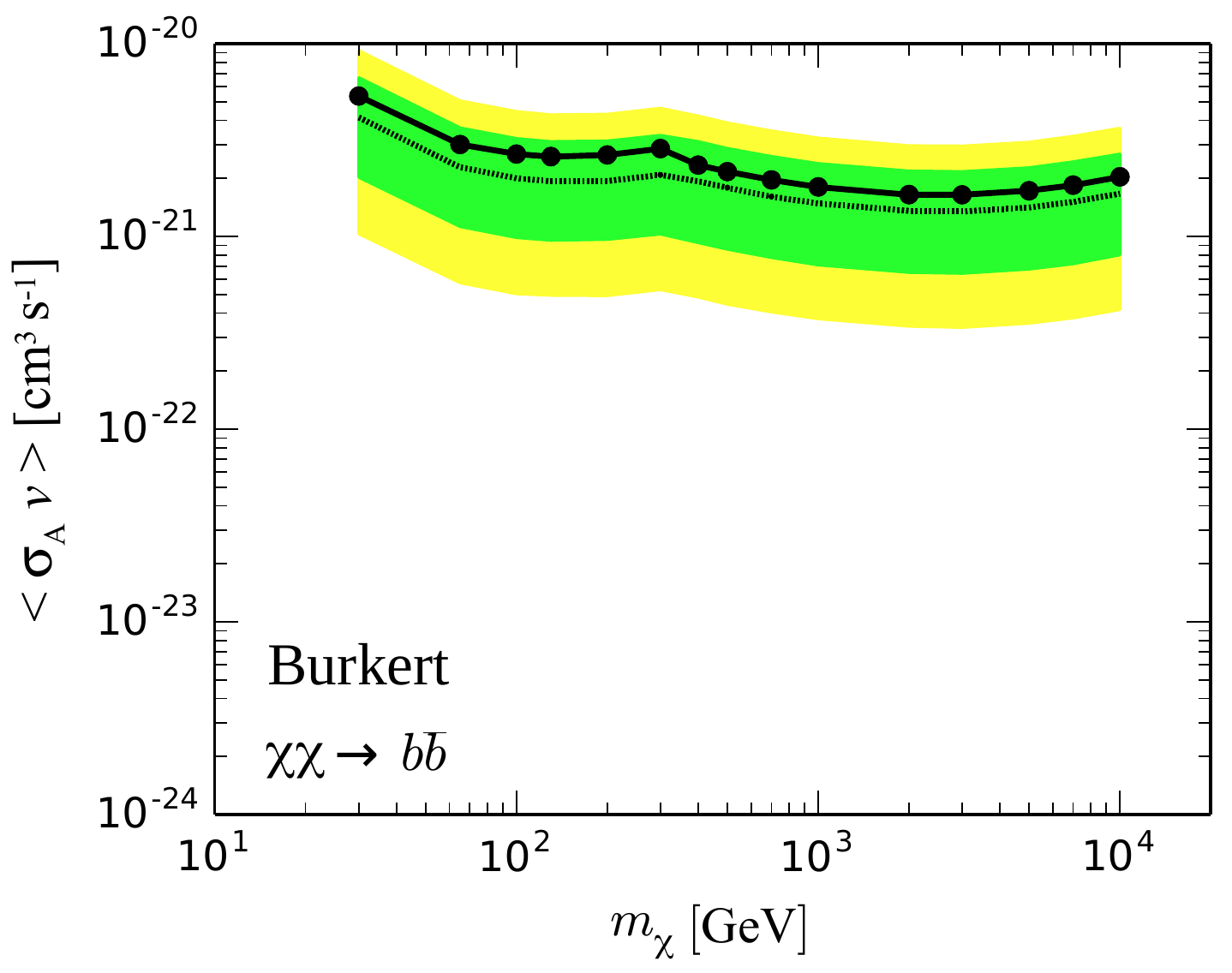}
\hfill
\includegraphics[width=0.45\linewidth,height=0.35\linewidth]{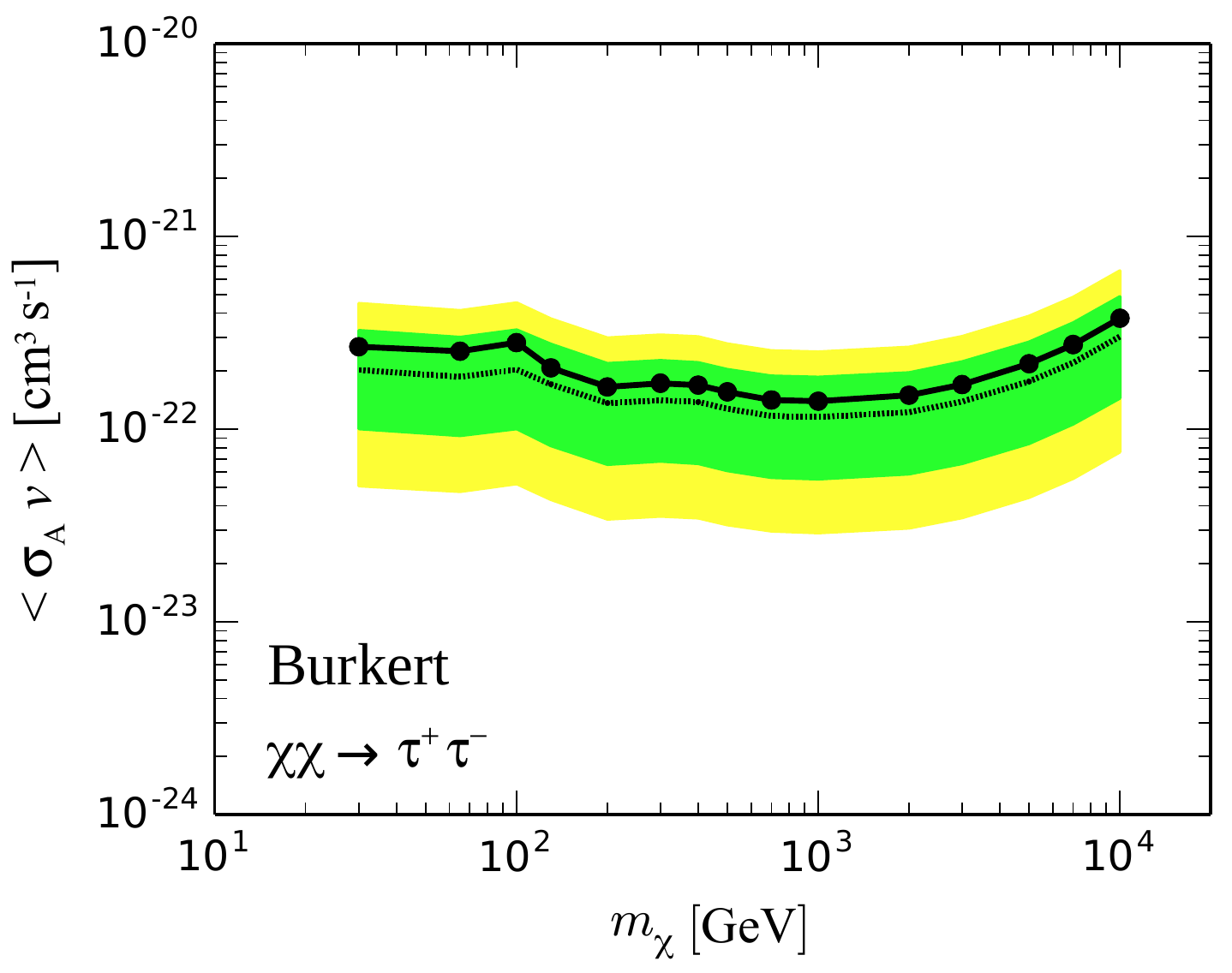}\\
\includegraphics[width=0.45\linewidth,height=0.35\linewidth]{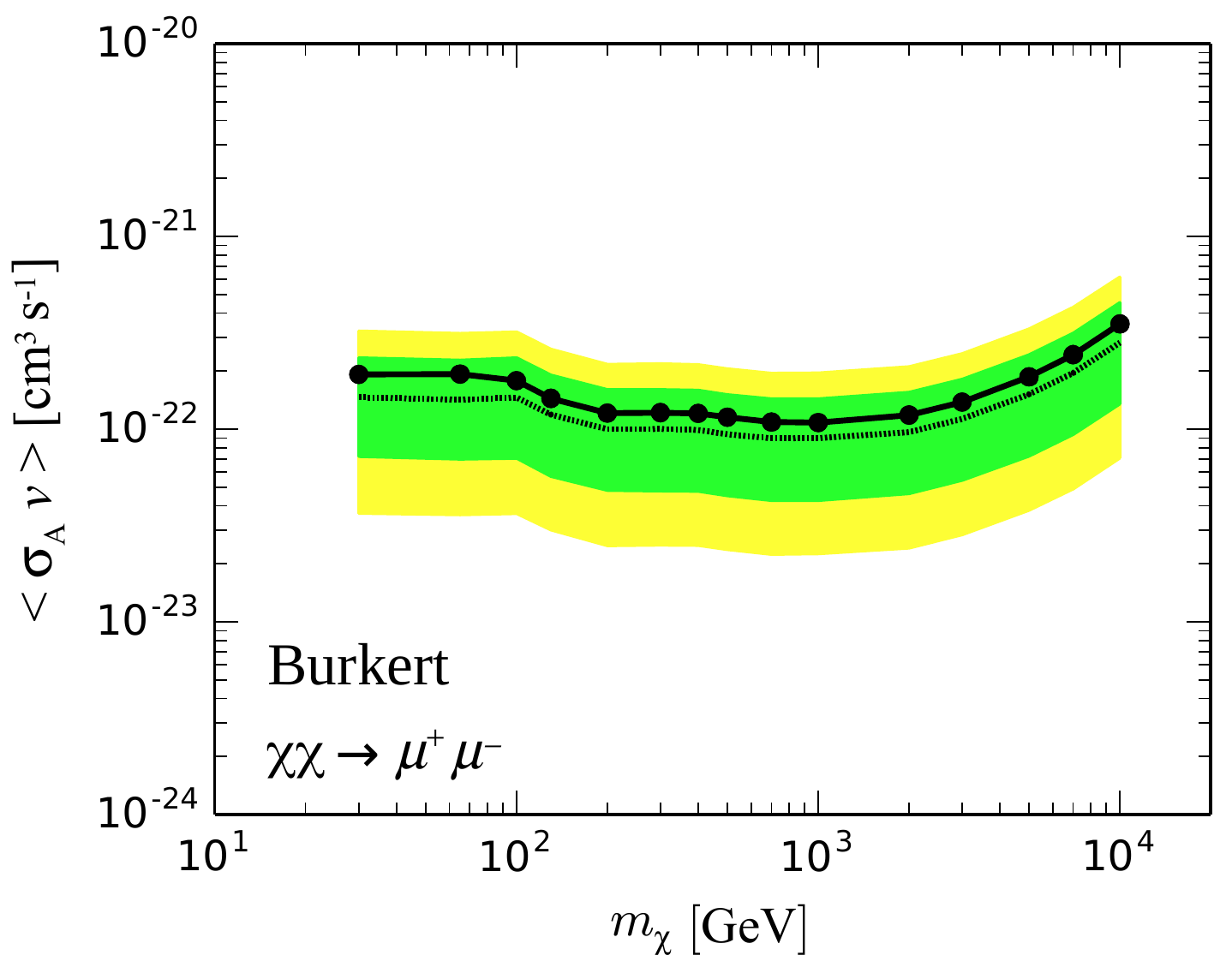}
\hfill
\includegraphics[width=0.45\linewidth,height=0.35\linewidth]{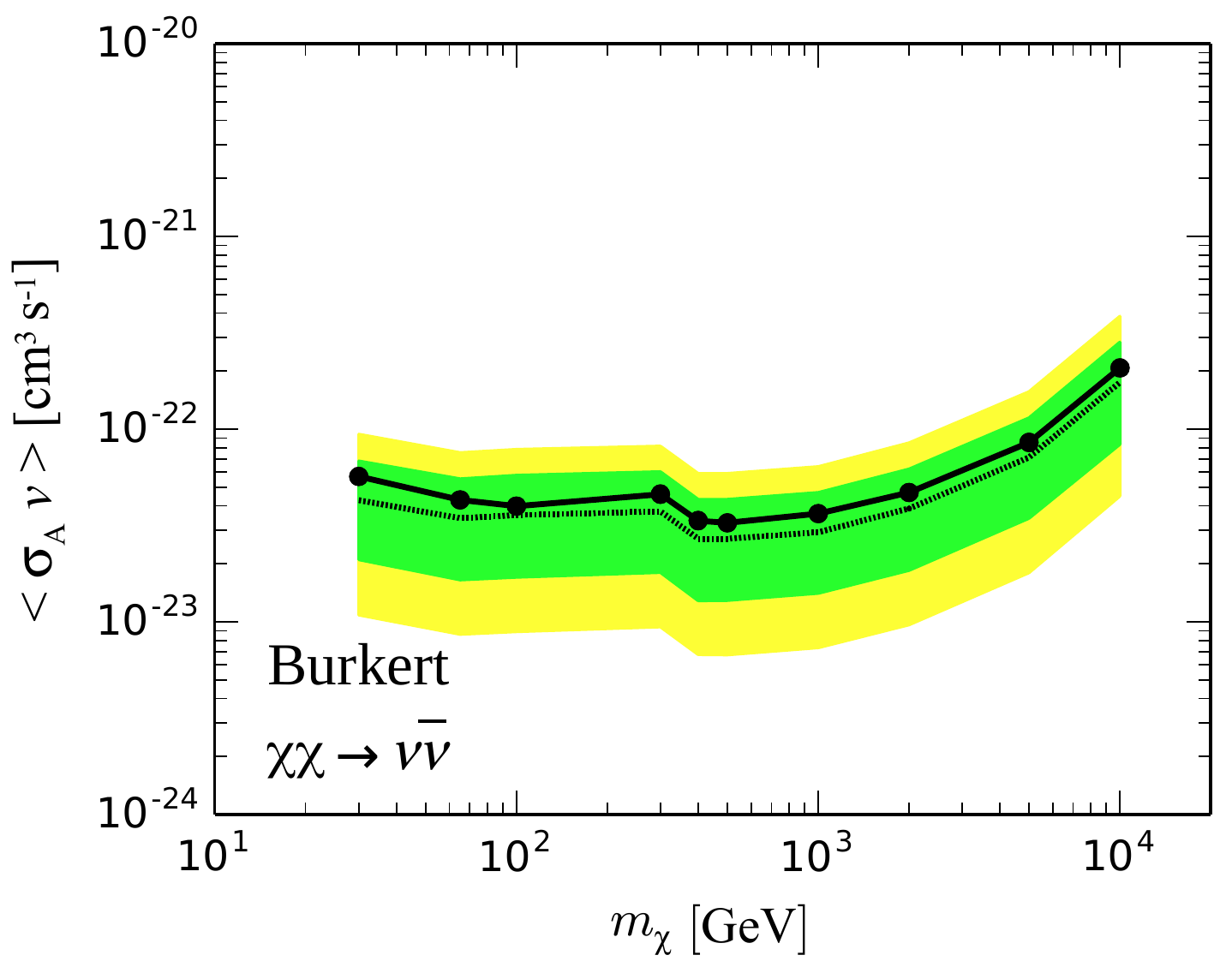}
\caption{Upper limits (90\% C.L., solid black line) on the velocity\hyp averaged WIMP self\hyp annihilation cross section, $\langle\sigma_{\mathrm{A}} \mathrm{v}\rangle$, for 
the Burkert halo model together with the corresponding sensitivities (dashed black line) and their 1$\sigma$ (green) and 2$\sigma$ (yellow) statistical uncertainties. The black 
dots represent the masses probed, while the black line in between is drawn to guide the eye. Each plot corresponds to a different annihilation channel as indicated in 
the legend. The local dark matter density used was $\rho_{\mathrm{local}}$ = 0.49 GeV/cm$^3$~\cite{Nesti:2013uwa}.}
\label{fig:UpperLimitsBurkert}
\end{figure*}

\section{Analysis method}\label{sec:analysis}
We use the distribution of the space angle $\psi$ between event directions and the Galactic Centre to construct a likelihood function and test the  
signal hypothesis (excess of events at small $\psi$ values) against the background\hyp only hypothesis (an event distribution isotropic in the sky). 
The signal and background hypotheses are represented by probability density functions  of the $\psi$ distributions, 
\begin{equation}
f(\psi \mid \mu) = \frac{\mu}{n_\mathrm{obs}} f_\mathrm{S}(\psi) + \left( 1 - \frac{\mu}{n_\mathrm{obs}} \right) f_\mathrm{B}(\psi \mid \mu),
\label{eq:combinedPDF}
\end{equation}
where the subscripts S and B denote signal and background respectively and $\mu$ is the number of signal events present among the total number of observed 
events, $n_\mathrm{obs}$. The angle  $\psi$ is allowed to be in the full range [0$^\circ$, 180$^\circ$], therefore covering the full sky, as shown in figure~\ref{fig:allPDFs}. 
This allows the analysis to be sensitive to the whole halo instead of just to the Galactic Centre. However, if the signal is allowed to come from anywhere in the halo, the 
background distribution, taken from data, is necessarily contaminated by a potential signal: thereby the dependence of $f_\mathrm{B}(\psi \mid \mu)$ on $\mu$ and not only 
on $\psi$. In particular the background distribution is constructed as 
\begin{equation}
f_\mathrm{B}(\psi \mid \mu) = \frac{\mu}{n_\mathrm{obs}} f_\mathrm{ss}(\psi) +  \left( 1 - \frac{\mu}{n_\mathrm{obs}} \right) f_\mathrm{sd}(\psi),
\label{eq:backgroundPDF}
\end{equation}
where $f_\mathrm{ss}$ and $f_\mathrm{sd}$ are the PDF of the scrambled arrival directions of signal simulation and data events respectively.

The likelihood that the data sample contains $\mu$ signal events is defined as
\begin{equation}
\mathcal{L}(\mu) = \prod_{i=1}^{n_\mathrm{obs}} f(\psi_i \mid \mu).
\label{eq:LikelihoodDefinition}
\end{equation}
where $n_\mathrm{obs}$ is the number of observed events and $f(\psi_i \mid \mu)$ is given in equation~(\ref{eq:combinedPDF}). We follow the method described 
in~\cite{Feldman:1997qc} to calculate a 90\% confidence level upper limit on $\mu$, $\mu_{\mathrm{90}}$, which gives an upper limit on the flux of neutrinos from 
the halo as defined in equation~(\ref{eq:NeutrinoFlux}). This limit can, in turn, be translated into a limit on $\langle \sigma_{\mathrm{A}} \mathrm{v} \rangle $ for any 
given WIMP mass, annihilation channel and halo profile. The final limits are shown in the next section, for the event selection that showed the best sensitivity 
in each case.

\section{Results and Conclusion}\label{sec:results}
 At final selection level, a total of 5892 (2178) events were observed in the full sky for the low\hyp energy (high\hyp energy) samples respectively.  
 Figure~\ref{fig:ang_dist} shows the angular distribution of the two event samples at final cut level. The distributions are compatible with 0 signal 
events for all WIMP masses and annihilation channels tested.  Tables~\ref{tab:bbSummary}, \ref{tab:ttSummary}, \ref{tab:mumuSummary} and~\ref{tab:nunuSummary} 
show the results for the best fit on the number of signal events, $\hat{\mu}$,  together with the 90\% upper limits on the number of signal events, $\mu_{90}$, 
and the corresponding limit on the thermally\hyp averaged WIMP annihilation cross section, $\langle \sigma_{\mathrm{A}} \mathrm{v} \rangle_{90}$. Corresponding 
quantities with a tilde denote median upper limits (i.e., sensitivities). Each table corresponds to a given benchmark annihilation channel and it shows 
different WIMP ma\-sses for the two halo models considered. The available statistics at final level in the case of direct annihilation of 700~GeV WIMPs 
to neutrinos using the Burkert profile were not sufficient to define an angular distribution which was smooth enough to perform the shape analysis, so we 
choose not to quote results for this mass and channel in table~\ref{tab:nunuSummary}. Figures~\ref{fig:UpperLimitsNFW} and~\ref{fig:UpperLimitsBurkert} show 
the results graphically for the NFW and Burkert dark matter profiles respectively. The plots show the 90\% C.L. upper limits (solid black line) on the 
velocity\hyp averaged WIMP self\hyp annihilation cross section, $\langle\sigma_{\mathrm{A}} \mathrm{v}\rangle$, together with the corresponding sensitivities 
(dashed black line) and the 1$\sigma$ (green) and 2$\sigma$ (yellow) statistical uncertainties.\par
 
\begin{figure}[t]
\includegraphics[width=\linewidth,height=0.85\linewidth]{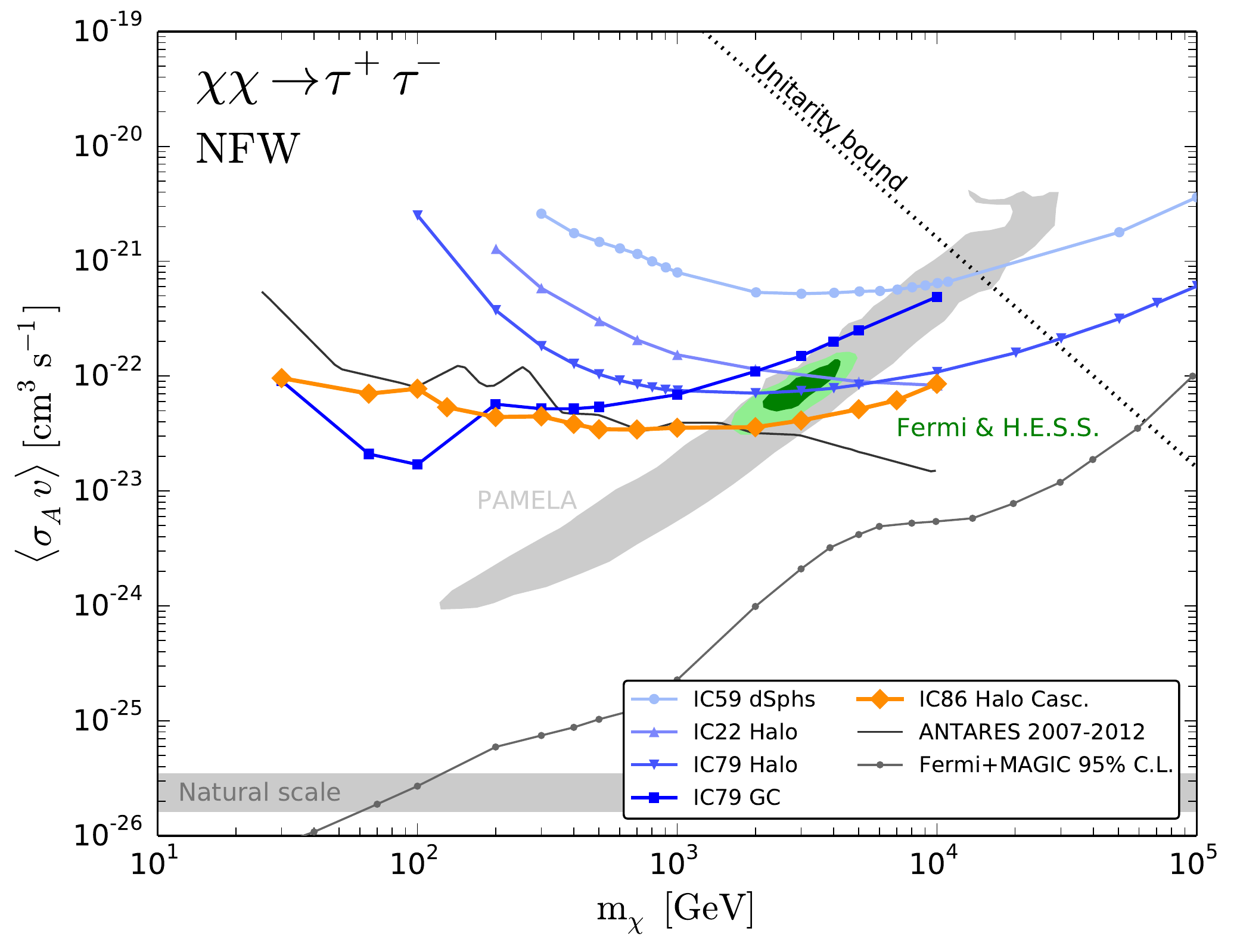}
\caption{Comparison of upper limits on $\langle \sigma_\mathrm{A} \mathrm{v} \rangle$ versus WIMP mass, for the annihilation channel 
$\chi\chi \rightarrow \tau^+\tau^-$. This work (IC86 Halo Casc.) is compared to \texttt{ANTARES} \cite{Adrian-Martinez:2015wey} and 
previous IceCube searches with different detector configurations~\cite{Abbasi:2011eq,Aartsen:2014hva,Aartsen:2013dxa,Aartsen:2015xej}. 
Also shown are the latest upper limits from gamma-ray searches obtained from the combination of \texttt{FermiLAT} and \texttt{MAGIC} results~\cite{Ahnen:2016qkx}. 
 The three shaded areas indicate allowed regions if the $e^++e^-$ flux excess seen by \texttt{Fermi-LAT}, \texttt{H.E.S.S.} and the positron 
excess seen by \texttt{PAMELA} ($3\sigma$ in dark green, $5\sigma$ in light green and gray area, respectively) would be interpreted as 
originating from dark-matter annihilations. The data for the shaded regions are taken from~\cite{Meade:2009iu}. The natural scale denotes 
the required value of $\langle \sigma_\mathrm{A} \mathrm{v} \rangle$ for a thermal-relic to constitute the dark matter~\cite{Steigman:2012nb}.}
\label{fig:comparison}
\end{figure}

In order to put the results of this analysis in perspective, figure~\ref{fig:comparison} shows a comparison with results from previous IceCube analyses 
and other experiments, for the $\tau\tau$ annihilation channel and the NFW profile. Also shown is the allowed area in the  
$(\langle\sigma_{\mathrm{A}} \mathrm{v}\rangle$, $m_{\chi})$ parameter space if the $e^++e^-$ flux excess seen by \texttt{Fermi-LAT} and \texttt{H.E.S.S.}  
and the positron excess seen by \texttt{PAMELA} are interpreted as originating from dark matter annihilations~\cite{Meade:2009iu}. There exist, however, 
conventional explanations based on local astrophysical sources~\cite{Blasi:2009hv,Hooper:2008kg} that, along with current limits on 
$\langle\sigma_{\mathrm{A}} \mathrm{v}\rangle$, disfavour such explanation. The figure shows that the analysis presented in this paper improves on previous 
IceCube analyses~\cite{Abbasi:2011eq,Aartsen:2014hva,Aartsen:2013dxa,Aartsen:2015xej} for WIMP masses above about 200~GeV, as well as on the \texttt{ANTARES}~\cite{Adrian-Martinez:2015wey} 
result for WIMP masses below $\sim$1~TeV. This demonstrates that particle cascades can be reconstructed with a good enough angular resolution in IceCube to make this 
channel competitive in searches for dark matter signals with neutrinos from the Galactic Centre and halo. 
 Even if Cherenkov telescopes and gamma-ray satellites can reach stricter bounds on $\langle \sigma_\mathrm{A} \mathrm{v} \rangle$ due to their better 
angular resolution and, depending on the source under consideration, low background, there is a much-needed complementarity in the field of dark matter searches, where neutrino 
telescopes can play a valuable role.

\begin{acknowledgements}
We acknowledge the support of the following institutions: U.S. National Science Foundation-Office of Polar Programs,
U.S. National Science Foundation-Physics Division,
University of Wisconsin Alumni Research Foundation,
the Grid Laboratory Of Wisconsin (GLOW) grid infrastructure at the University of Wisconsin - Madison, the Open Science Grid (OSG) grid infrastructure;
U.S. Department of Energy, and National Energy Research Scientific Computing Centre,
the Louisiana Optical Network Initiative (LONI) grid computing resources;
Natural Sciences and Engineering Research Council of Canada,
WestGrid and Compute/Calcul Canada;
Swedish Research Council,
Swedish Polar Research Secretariat,
Swedish National Infrastructure for Computing (SNIC),
and Knut and Alice Wallenberg Foundation, Sweden;
German Ministry for Education and Research (BMBF),
Deutsche Forschungsgemeinschaft (DFG),
Helmholtz Alliance for Astroparticle Physics (HAP),
Research Department of Plasmas with Complex Interactions (Bochum), Germany;
Fund for Scientific Research (FNRS-FWO),
FWO Odysseus programme,
Flanders Institute to encourage scientific and technological research in industry (IWT),
Belgian Federal Science Policy Office (Belspo);
University of Oxford, United Kingdom;
Marsden Fund, New Zealand;
Australian Research Council;
Japan Society for Promotion of Science (JSPS);
the Swiss National Science Foundation (SNSF), Switzerland;
National Research Foundation of Korea (NRF);
Villum Fonden, Danish National Research Foundation (DNRF), Denmark.
H.T. acknowledges support from the K. G. och Elisabeth Lennanders Stiftelse.

\end{acknowledgements}

\begin{table*}[ht]
\caption{Summary table of the results for the $\chi\chi \rightarrow b\bar{b}$ annihilation channel for both the Burkert and NFW halo profiles. The best fit for the number of signal events $\hat{\mu}$ is presented together with the upper limits $\mu_{90}$ and $\langle \sigma_{\mathrm{A}} \mathrm{v} \rangle_{90}$ along with their corresponding sensitivities $\tilde{\mu}_{90}$ and $\langle \widetilde{\sigma_{\mathrm{A}} \mathrm{v}} \rangle_{90}$. The values for each mass are presented for the event stream (LE or HE) with the best sensitivity.}
\begin{tabular*}{\textwidth}{@{\extracolsep{\fill}} |r  c|ccccc | ccccc |@{}}
\hline
  & & \multicolumn{5}{c|}{\textbf{Burkert profile}} & \multicolumn{5}{c|}{\textbf{NFW profile}} \\
  Mass & Selection & $\hat{\mu}$ & $\mu_{90}$ & $\tilde{\mu}_{90}$ & $\langle \sigma_A \mathrm{v} \rangle_{90}$ & $\langle \widetilde{\sigma_{\mathrm{A}} \mathrm{v}} \rangle_{90}$ &  $\hat{\mu}$ & $\mu_{90}$ & $\tilde{\mu}_{90}$ & $\langle \sigma_{\mathrm{A}} \mathrm{v} \rangle_{90}$ & $\langle \widetilde{\sigma_{\mathrm{A}} \mathrm{v}} \rangle_{90}$ \\[1mm]
  (GeV) & LE/HE & (\#)  & (\#)  & (\#)  & $\mathrm{(cm^3s^{-1})}$ & $\mathrm{(cm^3s^{-1})}$ &(\#)  & (\#)  & (\#)  & $\mathrm{(cm^3s^{-1})}$ & $\mathrm{(cm^3s^{-1})}$ \\
\hline
  30 & LE & 119 & 697 & 540 & $5.34\times 10^{-21}$ & $4.14\times 10^{-21}$ & 125 & 521 & 343 & $2.29\times 10^{-21}$ & $1.50\times 10^{-21}$\\
 65 & LE & 118 & 652 & 498 & $2.99\times 10^{-21}$ & $2.28\times 10^{-21}$ &  102 & 446 & 300 & $1.05\times 10^{-21}$ & $7.09\times 10^{-22}$\\
 100 & LE & 118 & 630 & 472 & $2.67\times 10^{-21}$ & $2.00\times 10^{-21}$ &  97.2 & 418 & 277 & $8.61\times 10^{-22}$ & $5.72\times 10^{-22}$\\[2mm]
  130 & LE & 118 & 614 & 458 & $2.59\times 10^{-21}$ & $1.92\times 10^{-21}$ &  93.8 & 401 & 265 & $8.14\times 10^{-22}$ & $5.39\times 10^{-22}$\\
  200 & LE & 118 & 593 & 435 & $2.64\times 10^{-21}$ & $1.94\times 10^{-21}$ &  87.5 & 373 & 246 & $7.94\times 10^{-22}$ & $5.23\times 10^{-22}$\\
  300 & LE & 116 & 574 & 419 & $2.85\times 10^{-21}$ & $2.08\times 10^{-21}$ &  83.7 & 357 & 233 & $8.41\times 10^{-22}$ & $5.51\times 10^{-22}$\\[2mm]
  400 & HE & 31.3 & 205 & 169 & $2.34\times 10^{-21}$ & $1.93\times 10^{-21}$ &  21.7 & 106 & 78.7 & $6.02\times 10^{-22}$ & $4.46\times 10^{-22}$\\
  500 & HE & 31.2 & 204 & 168 & $2.16\times 10^{-21}$ & $1.79\times 10^{-21}$ &  21.3 & 104 & 76.4 & $5.54\times 10^{-22}$ & $4.07\times 10^{-22}$\\
  700 & HE & 31.2 & 201 & 165 & $1.97\times 10^{-21}$ & $1.60\times 10^{-21}$ & 20.8 & 101 & 74.3 & $4.90\times 10^{-22}$ & $3.61\times 10^{-22}$\\[2mm]
  1000 & HE & 31.2 & 200 & 165 & $1.80\times 10^{-21}$ & $1.48\times 10^{-21}$ &  20.6 & 99.6 & 72.9 & $4.47\times 10^{-22}$ & $3.28\times 10^{-22}$\\
  2000 & HE & 30.5 & 199 & 164 & $1.64\times 10^{-21}$ & $1.35\times 10^{-21}$ &  20.4 & 98.0 & 71.6 & $4.16\times 10^{-22}$ & $3.05\times 10^{-22}$\\
  3000 & HE & 30.7 & 199 & 163 & $1.64\times 10^{-21}$ & $1.34\times 10^{-21}$ &  19.5 & 95.6 & 70.2 & $4.08\times 10^{-22}$ & $3.00\times 10^{-22}$\\[2mm]
  5000 & HE & 30.7 & 198 & 162 & $1.73\times 10^{-21}$ & $1.41\times 10^{-21}$ &  18.4 & 92.7 & 68.8 & $4.20\times 10^{-22}$ & $3.12\times 10^{-22}$\\
  7000 & HE & 30.8 & 197 & 161 & $1.83\times 10^{-21}$ & $1.51\times 10^{-21}$ & 17.8 & 91.1 & 67.8 & $4.45\times 10^{-22}$ & $3.30\times 10^{-22}$\\
  10000 & HE & 31.1 & 196 & 160 & $2.03\times 10^{-21}$ & $1.66\times 10^{-21}$ &  17.3 & 89.1 & 66.1 & $4.85\times 10^{-22}$ & $3.60\times 10^{-22}$\\[1.2mm]

\hline
\end{tabular*}
\label{tab:bbSummary}
\end{table*}%

\begin{table*}[ht]
\caption{Summary table of the results for the $\chi\chi \rightarrow \tau^+\tau^-$ annihilation channel for both the Burkert and NFW halo profiles. The best fit for the number of signal events $\hat{\mu}$ is presented together with the upper limits $\mu_{90}$ and $\langle \sigma_{\mathrm{A}} \mathrm{v} \rangle_{90}$ along with their corresponding sensitivities $\tilde{\mu}_{90}$ and $\langle \widetilde{\sigma_{\mathrm{A}} \mathrm{v}} \rangle_{90}$. The values for each mass are presented for the event stream (LE or HE) with the best sensitivity.}
\begin{tabular*}{\textwidth}{@{\extracolsep{\fill}} |r  c|ccccc | ccccc |@{}}
\hline
  & & \multicolumn{5}{c|}{\textbf{Burkert profile}} & \multicolumn{5}{c|}{\textbf{NFW profile}} \\
  Mass & Selection & $\hat{\mu}$ & $\mu_{90}$ & $\tilde{\mu}_{90}$ & $\langle \sigma_A \mathrm{v} \rangle_{90}$ & $\langle \widetilde{\sigma_{\mathrm{A}} \mathrm{v}} \rangle_{90}$ &  $\hat{\mu}$ & $\mu_{90}$ & $\tilde{\mu}_{90}$ & $\langle \sigma_{\mathrm{A}} \mathrm{v} \rangle_{90}$ & $\langle \widetilde{\sigma_{\mathrm{A}} \mathrm{v}} \rangle_{90}$ \\[1mm]
 (GeV) & LE/HE & (\#)  & (\#)  & (\#)  & $\mathrm{(cm^3s^{-1})}$ & $\mathrm{(cm^3s^{-1})}$ &(\#)  & (\#)  & (\#)  & $\mathrm{(cm^3s^{-1})}$ & $\mathrm{(cm^3s^{-1})}$ \\
\hline
  30 & LE & 118 & 651 & 494 & $2.67\times 10^{-22}$ & $2.03\times 10^{-22}$ & 96.1 & 443 & 305 & $9.61\times 10^{-23}$ & $6.62\times 10^{-23}$\\
  65 & LE & 118 & 594 & 437 & $2.54\times 10^{-22}$ & $1.86\times 10^{-22}$ &  89.5 & 378 & 249 & $7.03\times 10^{-23}$ & $4.62\times 10^{-23}$\\
  100 & LE & 116 & 554 & 402 & $2.80\times 10^{-22}$ & $2.03\times 10^{-22}$ & 78.3 & 334 & 219 & $7.77\times 10^{-23}$ & $5.08\times 10^{-23}$\\[2mm]
  130 & HE & 31.7 & 206 & 170 & $2.08\times 10^{-22}$ & $1.70\times 10^{-22}$ &  22.5 & 111 & 82.3 & $5.36\times 10^{-23}$ & $3.98\times 10^{-23}$\\
  200 & HE & 31.0 & 206 & 170 & $1.65\times 10^{-22}$ & $1.36\times 10^{-22}$ & 21.2 & 105 & 77.7 & $4.40\times 10^{-23}$ & $3.25\times 10^{-23}$\\
  300 & HE & 31.3 & 202 & 164 & $1.73\times 10^{-22}$ & $1.41\times 10^{-22}$ &  19.8 & 97.8 & 72.1 & $4.46\times 10^{-23}$ & $3.29\times 10^{-23}$\\[2mm]
  400 & HE & 31.6 & 200 & 163 & $1.69\times 10^{-22}$ & $1.37\times 10^{-22}$ &  19.5 & 95.3 & 69.9 & $3.83\times 10^{-23}$ & $2.81\times 10^{-23}$\\
  500 & HE & 31.9 & 199 & 163 & $1.56\times 10^{-22}$ & $1.27\times 10^{-22}$ &  19.3 & 94.5 & 69.5 & $3.44\times 10^{-23}$ & $2.53\times 10^{-23}$\\
  700 & HE & 29.8 & 199 & 164 & $1.41\times 10^{-22}$ & $1.17\times 10^{-22}$ &  20.3 & 97.0 & 70.8 & $3.43\times 10^{-23}$ & $2.50\times 10^{-23}$\\[2mm]
  1000 & HE & 29.7 & 198 & 164 & $1.39\times 10^{-22}$ & $1.15\times 10^{-22}$ &  20.3 & 95.8 & 69.5 & $3.55\times 10^{-23}$ & $2.58\times 10^{-23}$\\
  2000 & HE & 31.9 & 200 & 163 & $1.50\times 10^{-22}$ & $1.22\times 10^{-22}$ &  17.0 & 90.0 & 67.4 & $3.59\times 10^{-23}$ & $2.69\times 10^{-23}$\\
  3000 & HE & 31.2 & 197 & 161 & $1.70\times 10^{-22}$ & $1.39\times 10^{-22}$ &  16.4 & 87.7 & 65.7 & $4.10\times 10^{-23}$ & $3.07\times 10^{-23}$\\[2mm]
  5000 & HE & 32.6 & 195 & 158 & $2.19\times 10^{-22}$ & $1.76\times 10^{-22}$ &  16.2 & 84.3 & 62.4 & $5.15\times 10^{-23}$ & $3.81\times 10^{-23}$\\
  7000 & HE & 32.2 & 193 & 155 & $2.74\times 10^{-22}$ & $2.21\times 10^{-22}$ &  14.9 & 80.7 & 60.1 & $6.16\times 10^{-23}$ & $4.58\times 10^{-23}$\\
  10000 & HE & 31.7 & 191 & 153 & $3.76\times 10^{-22}$ & $3.02\times 10^{-22}$ & 14.5 & 80.1 & 60.0 & $8.57\times 10^{-23}$ & $6.43\times 10^{-23}$\\[1.2mm]
\hline
\end{tabular*}
\label{tab:ttSummary}
\end{table*}%

\begin{table*}[ht]
\caption{Summary table of the results for the $\chi\chi \rightarrow \mu^+\mu^-$ annihilation channel for both the Burkert and NFW halo profiles. The best fit for the number of signal events $\hat{\mu}$ is presented together with the upper limits $\mu_{90}$ and $\langle \sigma_{\mathrm{A}} \mathrm{v} \rangle_{90}$ along with their corresponding sensitivities $\tilde{\mu}_{90}$ and $\langle \widetilde{\sigma_{\mathrm{A}} \mathrm{v}} \rangle_{90}$. The values for each mass are presented for the event stream (LE or HE) with the best sensitivity.}
\begin{tabular*}{\textwidth}{@{\extracolsep{\fill}} |r  c|ccccc | ccccc |@{}}
\hline
  & & \multicolumn{5}{c|}{\textbf{Burkert profile}} & \multicolumn{5}{c|}{\textbf{NFW profile}} \\
  Mass & Selection & $\hat{\mu}$ & $\mu_{90}$ & $\tilde{\mu}_{90}$ & $\langle \sigma_A \mathrm{v} \rangle_{90}$ & $\langle \widetilde{\sigma_{\mathrm{A}} \mathrm{v}} \rangle_{90}$ &  $\hat{\mu}$ & $\mu_{90}$ & $\tilde{\mu}_{90}$ & $\langle \sigma_{\mathrm{A}} \mathrm{v} \rangle_{90}$ & $\langle \widetilde{\sigma_{\mathrm{A}} \mathrm{v}} \rangle_{90}$ \\[1mm]
 (GeV) & LE/HE & (\#)  & (\#)  & (\#)  & $\mathrm{(cm^3s^{-1})}$ & $\mathrm{(cm^3s^{-1})}$ &(\#)  & (\#)  & (\#)  & $\mathrm{(cm^3s^{-1})}$ & $\mathrm{(cm^3s^{-1})}$ \\
\hline
 30 & LE & 118 & 652 & 496 & $1.92\times 10^{-22}$ & $1.46\times 10^{-22}$ &  100 & 448 & 304 & $6.84\times 10^{-23}$ & $4.64\times 10^{-23}$\\
 65 & LE & 116 & 581 & 428 & $1.92\times 10^{-22}$ & $1.42\times 10^{-22}$ &  87.4 & 365 & 238 & $5.22\times 10^{-23}$ & $3.40\times 10^{-23}$\\
 100 & LE$^*$ & 114 & 535 & 383 & $2.22\times 10^{-22}$ & $1.59\times 10^{-22}$ &  24.3 & 116 & 86.5 & $4.40\times 10^{-23}$ & $3.27\times 10^{-23}$\\[2mm]
  130 & HE & 31.6 & 205 & 169 & $1.43\times 10^{-22}$ & $1.18\times 10^{-22}$ & 22.5 & 110 & 81.7 & $3.56\times 10^{-23}$ & $2.64\times 10^{-23}$\\
  200 & HE & 31.2 & 207 & 170 & $1.21\times 10^{-22}$ & $9.98\times 10^{-23}$ & 20.6 & 103 & 76.5 & $3.12\times 10^{-23}$ & $2.32\times 10^{-23}$\\
  300 & HE & 30.7 & 200 & 164 & $1.22\times 10^{-22}$ & $1.00\times 10^{-22}$ & 19.4 & 95.6 & 70.2 & $3.13\times 10^{-23}$ & $2.30\times 10^{-23}$\\[2mm]
  400 & HE & 30.9 & 196 & 161 & $1.20\times 10^{-22}$ & $9.89\times 10^{-23}$ &  19.2 & 92.9 & 67.7 & $2.84\times 10^{-23}$ & $2.06\times 10^{-23}$\\
  500 & HE & 31.4 & 196 & 160 & $1.15\times 10^{-22}$ & $9.41\times 10^{-23}$ &  19.3 & 92.6 & 67.4 & $2.65\times 10^{-23}$ & $1.93\times 10^{-23}$\\
  700 & HE & 30.0 & 197 & 162 & $1.08\times 10^{-22}$ & $8.96\times 10^{-23}$ &  20.1 & 95.7 & 69.9 & $2.70\times 10^{-23}$ & $1.96\times 10^{-23}$\\[2mm]
  1000 & HE & 29.0 & 196 & 163 & $1.08\times 10^{-22}$ & $8.99\times 10^{-23}$ &  20.6 & 96.3 & 70.1 & $2.87\times 10^{-23}$ & $2.09\times 10^{-23}$\\
  2000 & HE & 31.5 & 197 & 161 & $1.18\times 10^{-22}$ & $9.65\times 10^{-23}$ &  16.3 & 88.6 & 66.7 & $2.77\times 10^{-23}$ & $2.09\times 10^{-23}$\\
  3000 & HE & 30.6 & 195 & 159 & $1.37\times 10^{-22}$ & $1.13\times 10^{-22}$ & 15.1 & 85.1 & 64.7 & $3.19\times 10^{-23}$ & $2.43\times 10^{-23}$\\[2mm]
  5000 & HE & 32.1 & 193 & 157 & $1.86\times 10^{-22}$ & $1.51\times 10^{-22}$ &  14.5 & 80.7 & 60.8 & $4.13\times 10^{-23}$ & $3.11\times 10^{-23}$\\
  7000 & HE & 32.2 & 191 & 153 & $2.43\times 10^{-22}$ & $1.94\times 10^{-22}$ & 13.9 & 77.0 & 57.7 & $5.20\times 10^{-23}$ & $3.89\times 10^{-23}$\\
  10000 & HE & 32.1 & 189 & 151 & $3.50\times 10^{-22}$ & $2.80\times 10^{-22}$ &  13.2 & 75.3 & 56.7 & $7.63\times 10^{-23}$ & $5.74\times 10^{-23}$\\[1.2mm]
\hline
\multicolumn{12}{l}{($^*$) HE event selection for the NFW profile} \\
\end{tabular*}
\label{tab:mumuSummary}
\end{table*}%

\begin{table*}[ht]
\caption{Summary table of the results for the $\chi\chi \rightarrow \nu\bar{\nu}$ annihilation channel for both the Burkert and NFW halo profiles. The best fit for the number of signal events $\hat{\mu}$ is presented together with the upper limits $\mu_{90}$ and $\langle \sigma_{\mathrm{A}} \mathrm{v} \rangle_{90}$ along with their corresponding sensitivities $\tilde{\mu}_{90}$ and $\langle \widetilde{\sigma_{\mathrm{A}} \mathrm{v}} \rangle_{90}$. The values for each mass are presented for the event stream (LE or HE) with the best sensitivity. The available statistics at final level for the 700 GeV WIMP sample under the Burkert profile were not sufficient to define an angular distribution which was smooth enough to perform the shape analysis, and we chose not to quote results for this case.}
\begin{tabular*}{\textwidth}{@{\extracolsep{\fill}} |r  c|ccccc | ccccc |@{}}
\hline
  & & \multicolumn{5}{c|}{\textbf{Burkert profile}} & \multicolumn{5}{c|}{\textbf{NFW profile}} \\
  Mass & Selection & $\hat{\mu}$ & $\mu_{90}$ & $\tilde{\mu}_{90}$ & $\langle \sigma_A \mathrm{v} \rangle_{90}$ & $\langle \widetilde{\sigma_{\mathrm{A}} \mathrm{v}} \rangle_{90}$ &  $\hat{\mu}$ & $\mu_{90}$ & $\tilde{\mu}_{90}$ & $\langle \sigma_{\mathrm{A}} \mathrm{v} \rangle_{90}$ & $\langle \widetilde{\sigma_{\mathrm{A}} \mathrm{v}} \rangle_{90}$ \\[1mm]
 (GeV) & LE/HE & (\#)  & (\#)  & (\#)  & $\mathrm{(cm^3s^{-1})}$ & $\mathrm{(cm^3s^{-1})}$ &(\#)  & (\#)  & (\#)  & $\mathrm{(cm^3s^{-1})}$ & $\mathrm{(cm^3s^{-1})}$ \\
\hline
  30 & LE & 109 & 585 & 441 & $5.67\times 10^{-23}$ & $4.28\times 10^{-23}$ &  75.6 & 354 & 242 & $1.59\times 10^{-23}$ & $1.08\times 10^{-23}$\\
  65 & HE & 34.6 & 200 & 161 & $4.29\times 10^{-23}$ & $3.46\times 10^{-23}$ &  25.0 & 112 & 80.9 & $1.35\times 10^{-23}$ & $9.80\times 10^{-24}$\\
  100 & HE & 16.7 & 181 & 163 & $3.99\times 10^{-23}$ & $3.58\times 10^{-23}$ & 18.7 & 88.3 & 61.9 & $1.20\times 10^{-23}$ & $8.43\times 10^{-24}$\\[2mm]
  300 & HE & 31.5 & 194 & 158 & $4.59\times 10^{-23}$ & $3.73\times 10^{-23}$ &  16.7 & 93.7 & 72.3 & $1.13\times 10^{-23}$ & $8.70\times 10^{-24}$\\
  400 & HE & 35.0 & 201 & 162 & $3.35\times 10^{-23}$ & $2.69\times 10^{-23}$ &  16.7 & 82.3 & 60.4 & $7.47\times 10^{-24}$ & $5.48\times 10^{-24}$\\
  500 & HE & 30.1 & 198 & 164 & $3.26\times 10^{-23}$ & $2.69\times 10^{-23}$ & 28.4 & 105 & 68.1 & $9.82\times 10^{-24}$ & $6.37\times 10^{-24}$\\[2mm]
  700 & HE  & -- & -- & -- & -- & -- &  24.6 & 104 & 71.5 & $1.05\times 10^{-23}$ & $7.25\times 10^{-24}$\\
  1000 & HE & 34.6 & 205 & 164 & $3.65\times 10^{-23}$ & $2.91\times 10^{-23}$ &  14.7 & 82.9 & 62.6 & $7.83\times 10^{-24}$ & $5.91\times 10^{-24}$\\
  2000 & HE & 28.4 & 188 & 155 & $4.69\times 10^{-23}$ & $3.87\times 10^{-23}$ & 16.7 & 85.7 & 62.9 & $1.43\times 10^{-23}$ & $1.05\times 10^{-23}$\\[2mm]
  5000 & HE & 25.0 & 177 & 148 & $8.53\times 10^{-23}$ & $7.15\times 10^{-23}$ &  12.6 & 69.4 & 52.4 & $1.38\times 10^{-23}$ & $1.05\times 10^{-23}$\\
  10000 & HE & 19.7 & 162 & 137 & $2.08\times 10^{-22}$ & $1.75\times 10^{-22}$ &  3.5 & 59.9 & 48.3 & $4.50\times 10^{-23}$ & $3.64\times 10^{-23}$\\[1.2mm]
\hline
\end{tabular*}
\label{tab:nunuSummary}
\end{table*}%


\newpage

\begin{thebibliography}{10}
\providecommand{\url}[1]{{#1}}
\providecommand{\urlprefix}{URL }
\expandafter\ifx\csname urlstyle\endcsname\relax
  \providecommand{\doi}[1]{DOI \discretionary{}{}{}#1}\else
  \providecommand{\doi}{DOI \discretionary{}{}{}\begingroup
  \urlstyle{rm}\Url}\fi

\bibitem{Lukovic:2014vma}
  V.~Lukovic, P.~Cabella and N.~Vittorio,  Int.\ J.\ Mod.\ Phys.\ A {\bf 29}, 1443001 (2014).

\bibitem{Jungman:1995df}
G.~Jungman, M.~Kamionkowski and K.~Griest, Phys. Rept. \textbf{267}, 195 (1996).

\bibitem{Feng:2010gw}
J.~L. Feng, Ann. Rev. Astron. Astrophys. \textbf{48}, 495 (2010).

\bibitem{Bergstrom:2012fi}
L.~Bergstr\"om, Annalen Phys. \textbf{524}, 479 (2012).

\bibitem{Bergstrom:1997fj}
L.~Bergstr\"om, P.~Ullio and J.~H. Buckley, Astropart. Phys. \textbf{9}, 137 (1998).

\bibitem{Yuksel:2007ac}
H.~Yuksel, S.~Horiuchi, J.~F. Beacom and S.~Ando, Phys. Rev. \textbf{D76}, 123506  (2007).

\bibitem{Kravtsov:1997dp}
  A.~V.~Kravtsov, A.~A.~Klypin, J.~S.~Bullock and J.~R.~Primack,  Astrophys.\ J.\  {\bf 502}, 48, (1998).

\bibitem{Burkert:1995yz}
A.~Burkert, Astrophys. J. \textbf{171}, 175 (1996).

\bibitem{Navarro:1995iw}
J.~F. Navarro, C.~S. Frenk and S.~D. White, Astrophys. J. \textbf{462}, 563 (1996).

\bibitem{Diemand:2009bm}
  J.~Diemand and B.~Moore, Adv.\ Sci.\ Lett.\  {\bf 4}, 297 (2011).
 
\bibitem{Ruffini:2014zfa} 
  R.~Ruffini, C.~R.~Arg\"uelles and J.~A.~Rueda,  Mon.\ Not.\ Roy.\ Astron.\ Soc.\  {\bf 451}, no. 1, 622 (2015).

\bibitem{deBlok:2009sp}
W.~J.~G.~de Blok,   Adv.\ Astron.\  {\bf 2010}, 789293, (2010).

\bibitem{Salucci:2000ps}
P.~Salucci and A.~Burkert, Astrophys. J. \textbf{537}, L9 (2000).

\bibitem{Nesti:2013uwa}
F.~Nesti and P.~Salucci, JCAP {\bf 1307} (2013) 016.

\bibitem{Adriani:2008zr}
O.~Adriani et~al., Nature {\bf 458}, 607 (2009).

\bibitem{Abdo:2009zk}
A.~A.~Abdo et~al., Phys. Rev. Lett. {\bf 102}, 181101 (2009).

\bibitem{Chang:2008aa}
J.~Chang et~al., Nature {\bf 456}, 362 (2008).

\bibitem{Cirelli:2008pk}
M.~Cirelli, M.~Kadastik, M.~Raidal and A.~Strumia, Nucl. Phys. B{\bf 813}, 1 (2009). Addendum Nucl. Phys. B{\bf 873}, 530 (2013).

\bibitem{Donato:2008jk}
F.~Donato, D.~Maurin, P.~Brun, T.~Delahaye and P.~Salati, Phys.\ Rev.\ Lett.\  {\bf 102},  071301 (2009).

\bibitem{Barger:2008su}
V.~Barger, W.~Y.~Keung, D.~Marfatia and G.~Shaughnessy,  Phys. Lett. B {\bf 672}, 141 (2009).

\bibitem{Bergstrom:2009fa}
L.~Bergstr\"om, J.~Edsj\"o and  G.~Zaharijas,  Phys.\ Rev.\ Lett.\  {\bf 103},  031103 (2009).

\bibitem{Cholis:2008hb}
I.~Cholis, L.~Goodenough, D.~Hooper, M.~Simet and N.~Weiner,  Phys.\ Rev.\ D {\bf 80},  123511 (2009).

\bibitem{Mandal:2009yk}
S.~K. Mandal, M.~R. Buckley, K.~Freese, D.~Spolyar, H.~Murayama, Phys. Rev. \textbf{D81}, 043508 (2010).

\bibitem{Halzen:2010yj}
F.~Halzen and S.~R. Klein, Rev. Sci. Instrum. \textbf{81}, 081101 (2010).

\bibitem{IceCube:2012nn}
R.~Abbasi, et~al., Nucl. Instrum. Meth. \textbf{A700}, 188 (2013).

\bibitem{Collaboration:2011ym}
R.~Abbasi, et~al., Astropart. Phys. \textbf{35}, 615 (2012).

\bibitem{Sjostrand:2007gs}
T.~Sjostrand, S.~Mrenna and P.~Z. Skands, Comput. Phys. Commun. \textbf{178}, 852 (2008).

\bibitem{Barger:2007xf}
V.~Barger, W-Y. Keung, G. Shaughnessy and A. Tregre, Phys. Rev. \textbf{D76}, 095008 (2007).

\bibitem{Heck:1998vt}
D.~Heck, G.~Schatz, T.~Thouw, J.~Knapp and J.~N. Capdevielle, FZKA Technical Report.  (1998)

\bibitem{Andreopoulos:2009rq}
C.~Andreopoulos, et~al., Nucl. Instrum. Meth. \textbf{A614}, 87 (2010).

\bibitem{Gazizov:2004va}
A.~Gazizov and  M.~P. Kowalski, Comput. Phys. Commun. \textbf{172}, 203 (2005).

\bibitem{Honda:2006qj}
M.~Honda, T.~Kajita, K.~Kasahara, S.~Midorikawa and T.~Sanuki, Phys. Rev. \textbf{D75}, 043006 (2007).

\bibitem{Chirkin:2013tma}
D.~Chirkin, Nucl. Instrum. Meth. \textbf{A725}, 141 (2013).

\bibitem{Aartsen:2013bfa}
M.~G. Aartsen, et~al., Nucl. Instrum. Meth. \textbf{A736}, 143 (2014).

\bibitem{Ahrens:2003fg}
J.~Ahrens, et~al., Nucl. Instrum. Meth. \textbf{A524}, 169 (2004).

\bibitem{Hocker:2007ht}
A.~Hocker, et~al., PoS \textbf{ACAT}, 040 (2007).

\bibitem{Aartsen:2013rt}
M.~G. Aartsen, et~al., Nucl. Instrum. Meth. \textbf{A711}, 73 (2013).

\bibitem{Dima:13aa}
M.~G. Aartsen, et~al., in \emph{Proceedings of the 33rd Cosmic Ray Conference,
  ICRC2013. Rio de Janeiro, Brazil.} (2013), 0580.

\bibitem{Feldman:1997qc}
G.~J. Feldman and R.~D. Cousins, Phys. Rev. \textbf{D57}, 3873 (1998).

\bibitem{Meade:2009iu}
P.~Meade, M.~Papucci, A.~Strumia and T.~Volansky, Nucl. Phys. \textbf{B831}, 178 (2010).

\bibitem{Blasi:2009hv}
P.~Blasi, Phys. Rev. Lett. \textbf{103}, 051104 (2009).

\bibitem{Hooper:2008kg}
D.~Hooper, P.~Blasi and P.~D. Serpico, JCAP \textbf{0901}, 025 (2009).

\bibitem{Abbasi:2011eq}
R.~Abbasi, et~al., Phys. Rev. \textbf{D84}, 022004 (2011).

\bibitem{Aartsen:2014hva}
M.~G. Aartsen, et~al., Eur. Phys. J. \textbf{C75}(1), 20 (2015).

\bibitem{Aartsen:2013dxa}
M.~G. Aartsen, et~al., Phys. Rev. \textbf{D88}, 122001 (2013).

\bibitem{Aartsen:2015xej}
M.~G. Aartsen, et~al., Eur. Phys. J. \textbf{C75}(10), 492 (2015).

\bibitem{Adrian-Martinez:2015wey}
S.~Adrian-Martinez, et~al., JCAP \textbf{1510}(10), 068  (2015).

\bibitem{Ahnen:2016qkx}
M.~L. Ahnen, et~al., JCAP \textbf{1602}(02), 039 (2016).

\bibitem{Steigman:2012nb}
G.~Steigman, B.~Dasgupta and J.~F. Beacom, Phys. Rev. \textbf{D86}, 023506 (2012).

\end{thebibliography}
\end{document}